\documentclass[%
 aip,
 jmp,%
 amsmath,amssymb,
 reprint,%
]{revtex4-1}

\pdfoutput=1

\usepackage[T1]{fontenc}
\usepackage[latin9]{inputenc}
\usepackage{geometry}
\usepackage{textcomp}
\usepackage{amstext}
\usepackage{amsmath}
\usepackage{graphicx}
\usepackage[english]{babel}
\usepackage{color}
\usepackage{multirow}

\begin{document}

% Text Color Macros for Comments
\newcommand{\tcb}{\textcolor{blue}}
\newcommand{\tcr}{\textcolor{red}}
\newcommand{\tcg}{\textcolor{green}}

% Good Tilde Macro
\newcommand{\tildegood}{{\raise.17ex\hbox{$\scriptstyle\sim$}}}

%==================================================================================================
%  Title, Authors, and Abstract
%==================================================================================================

%--------------------------------------------------------------------------------------------------
%  Title
%--------------------------------------------------------------------------------------------------
\title[aLIGO Input Optics]{The Advanced LIGO Input Optics}

%--------------------------------------------------------------------------------------------------
%  Authors
%--------------------------------------------------------------------------------------------------
\author{Chris L. Mueller}
\email{cmueller@phys.ufl.edu}

\author{Muzammil A. Arain}
\altaffiliation[Present Address: ]{KLA-Tencor, Milpitas, CA 95035, USA}

\author{Giacomo Ciani}

\affiliation{University of Florida, Gainesville, FL 32611, USA}

\author{Ryan. T. DeRosa}

\author{Anamaria Effler}

\affiliation{Louisiana State University, Baton Rough, LA 70803, USA}

\author{David Feldbaum}

\affiliation{University of Florida, Gainesville, FL 32611, USA}

\author{Valery V. Frolov}
\affiliation{LIGO Livingston Observatory, Livingston, LA 70754, USA}

\author{Paul Fulda}

\author{Joseph Gleason}
\altaffiliation[Present Address: ]{\tcr{Get Current Address 1}}

\author{Matthew Heintze}

\affiliation{University of Florida, Gainesville, FL 32611, USA}

\author{Eleanor J. King}
\affiliation{University of Adelaide, Adelaide, SA 5005, Australia}

\author{Keiko Kokeyama}
\affiliation{Louisiana State University, Baton Rough, LA 70803, USA}

\author{William Z. Korth}
\affiliation{LIGO, California Institute of Technology, Pasadena, CA 91125, USA}

\author{Rodica M. Martin}
\altaffiliation[Present Address: ]{\tcr{Get Current Address 2}}

\affiliation{University of Florida, Gainesville, FL 32611, USA}

\author{Adam Mullavey}
\affiliation{LIGO Livingston Observatory, Livingston, LA 70754, USA}

\author{Jan Poeld}
\affiliation{Max-Planck-Institut f\"{u}r Gravitationsphysik, 30167 Hannover, Germany}

\author{Volker Quetschke}
\affiliation{University of Texas at Brownsville, Brownsville, TX 78520, USA}

\author{David H. Reitze}
\altaffiliation[Present Address: ]{LIGO Laboratory, California Institute of Technology, Pasadena, 
	CA 91125, USA}

\author{David B. Tanner}

\author{Luke F. Williams}

\author{Guido Mueller}

\affiliation{University of Florida, Gainesville, FL 32611, USA}

%--------------------------------------------------------------------------------------------------
%  Abstract
%--------------------------------------------------------------------------------------------------
\begin{abstract}
	The Advanced LIGO gravitational wave detectors are nearing their design sensitivity and should 
	begin taking meaningful astrophysical data in the fall of 2015.  
	These resonant optical interferometers will have unprecedented sensitivity to the strains caused 
	by passing gravitational waves.  
	The input optics play a significant part in allowing these devices to reach such sensitivities.  
	
	Residing between the pre-stabilized laser and the main interferometer, the input optics is 
	tasked with preparing the laser beam for interferometry at the sub-attometer level while 
	operating at continuous wave input power levels ranging from 100 mW to 150 W.  
	These extreme operating conditions required every major component to be custom designed. 
	These designs draw heavily on the experience and understanding gained during the operation of  
	Initial LIGO and Enhanced LIGO.  
	In this article we report on how the components of the input optics were designed to meet their 
	stringent requirements and present measurements showing how well they have lived up to their 
	design.  
\end{abstract}

%--------------------------------------------------------------------------------------------------
%  Other Title Related Things
%--------------------------------------------------------------------------------------------------
% Date
\date{\today}

% Make the title portion
\maketitle

%==================================================================================================
%  Introduction 
%==================================================================================================
\section{Introduction}

A worldwide effort to directly detect gravitational radiation in the
$10\,\mbox{Hz}$ to a few kHz frequency range with large scale laser
interferometers has been underway for the past two decades. In the
United States the Laser Interferometer Gravitational-Wave Observatories
(LIGO) in Livingston, LA (LLO) and in Hanford, WA, (LHO) have been operating since the early 2000\textquoteright s.
Initial and enhanced LIGO produced several significant upper limits,
but did not have the sensitivity to make the first direct detection
of gravitational waves. During this time of operation a significant
amount of effort was invested by the LIGO Scientific Collaboration
to research and design Advanced LIGO (aLIGO), the first major upgrade of initial
LIGO. In 2011 the initial LIGO detectors were decommissioned and
installation of these upgrades started. The installation was completed in 2014 and 
the commissioning phase has begun for many of the
upgraded subsystems at the LIGO observatories. This paper focuses
on the input optics of aLIGO. 

\begin{figure}[t]
	\centering
	\includegraphics[width=\linewidth]{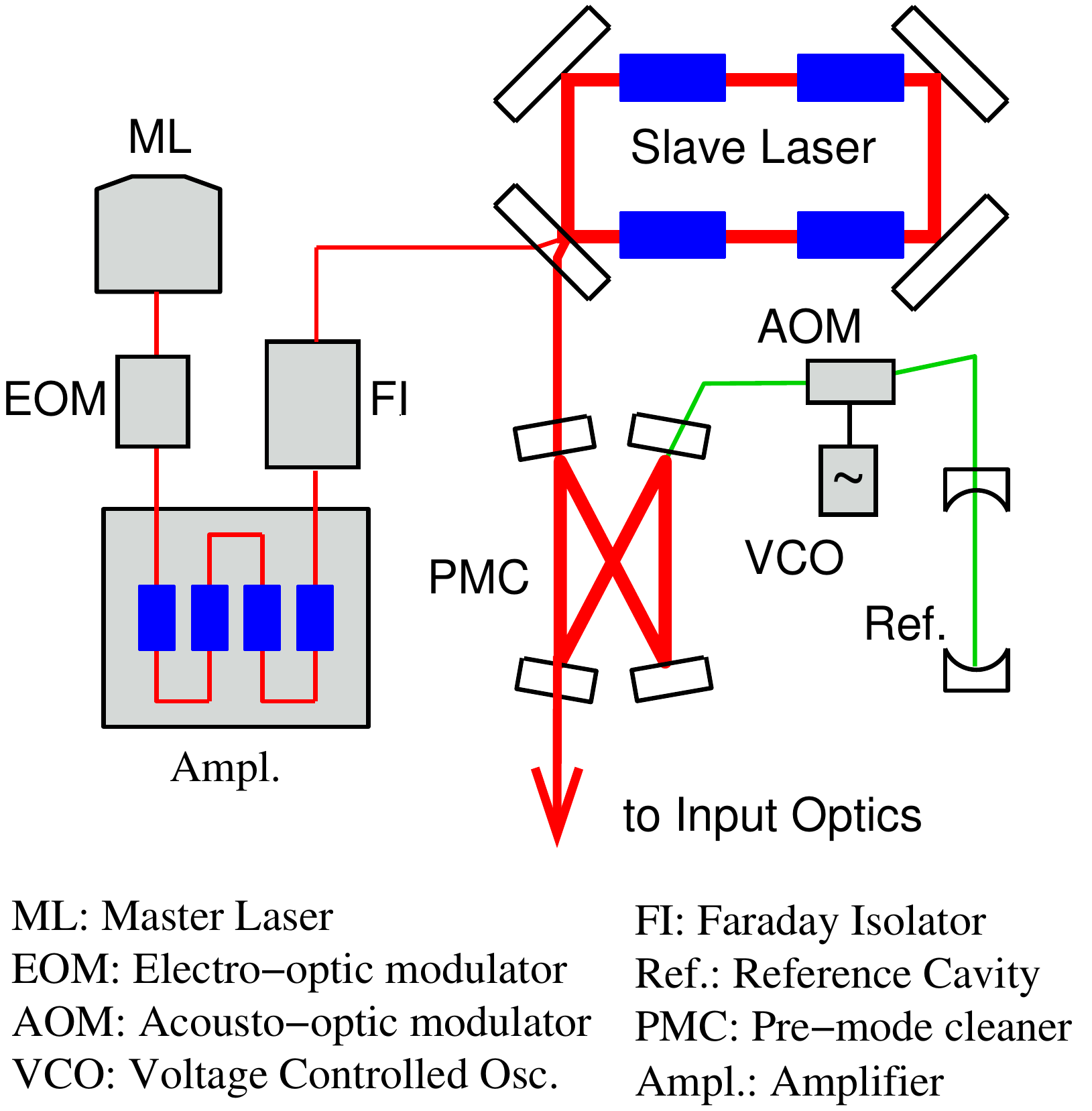}
	\caption{Sketch of the pre-stabilized laser (PSL) system.  
		Red: Main beam, Green: Pick-off beam.  
		The figure shows the low power master laser, the phase-correcting EOM, the amplifier stage, a 
		high power Faraday isolator, and the high power slave laser.  
		The pre-mode cleaner suppresses higher order spatial modes of the laser beam.  
		The VCO drives the AOM which shifts the frequency of the pick-off beam to stabilize the 
		sideband and with it the frequency of the laser to the reference cavity. }
	\label{fig:PSL}
\end{figure}

The main task of the input optics (IO) subsystem is to take the laser
beam from the pre-stabilized laser system\cite{kwee_stabilized_2012} (PSL) and prepare and inject
it into the main interferometer (IFO). The PSL consists of a master
laser, an amplifier stage, and a 200W slave laser which is injection-locked to the amplified master laser. The 200W output beam is filtered
by a short optical ring cavity, the pre-mode cleaner, before it is
turned over to the IO (see Figure \ref{fig:PSL}). The
PSL pre-stabilizes the laser frequency to a fixed spacer reference
cavity using a tunable sideband locking technique. The PSL also provides
interfaces to further stabilize its frequency and power. 

%^^^^^^^^^^^^^^^^^^^^^^^^^^^^^^^^^^^^^^^^^^^^^^^^^^^^^^^^^^^^^^^^^^^^^^^^^^^^^^^^^^^^^^^^^^^^^^^^^^
%^^^^^^^^^^^^^^^^^^^^^^^^^^^^^^^^^^^^^^^^^^^^^^^^^^^^^^^^^^^^^^^^^^^^^^^^^^^^^^^^^^^^^^^^^^^^^^^^^^		

%^^^^^^^^^^^^^^^^^^^^^^^^^^^^^^^^^^^^^^^^^^^^^^^^^^^^^^^^^^^^^^^^^^^^^^^^^^^^^^^^^^^^^^^^^^^^^^^^^^
\begin{figure}[t]
	\centering
	\includegraphics[width=\linewidth]{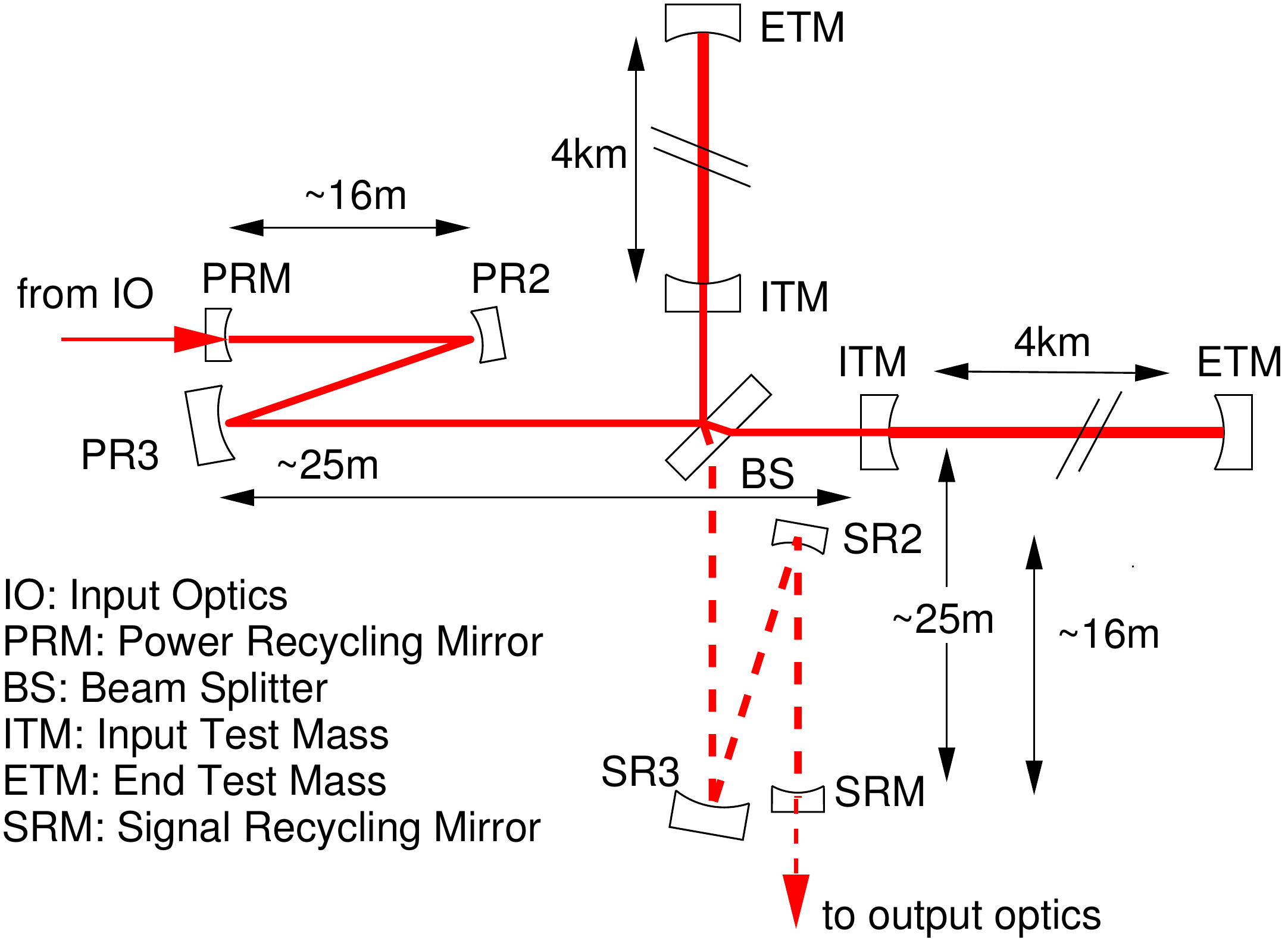}
	\caption{Sketch of the main interferometer which consists of two 4~km arm cavities, the beam 
		splitter, and the folded 55~m long power and signal recycling cavities. 
		The input optics is located between	this system and the PSL shown in Figure~\ref{fig:PSL}. }
	\label{fig:IFO}
\end{figure}
%^^^^^^^^^^^^^^^^^^^^^^^^^^^^^^^^^^^^^^^^^^^^^^^^^^^^^^^^^^^^^^^^^^^^^^^^^^^^^^^^^^^^^^^^^^^^^^^^^^

The IFO is a dual-recycled, cavity-enhanced Michelson interferometer\cite{harry_advanced_2010}
as sketched in Figure \ref{fig:IFO}. The field
enters the 55~m folded power recycling cavity (PRC) through
the power recycling mirror (PRM). Two additional mirrors (PR2, PR3) within the PRC form a fairly
fast telescope to increase the beam size from \tildegood2~mm
to \tildegood50~mm (Gaussian beam radius) before the large beam
is split at the beam splitter (BS) and injected into the two 4~km 
arm cavities formed by the input and end test masses. The reflected
fields recombine at the BS and send most of the light back to the
PRM where it constructively interferes with the injected field\cite{meers_recycling_1988}. This
leads to a power enhancement inside the power recycling cavity and
provides additional spatial, frequency, and amplitude filtering of
the laser beam. The second output of the BS sends light into the 55~m
long folded signal recycling cavity\cite{mizuno_resonant_1993} (SRC) which also consists of a
beam reducing telescope (SR2, SR3) and the semi-transparent signal
recycling mirror (SRM). 

This paper is organized as follows: Section \ref{sec:overview} gives an overview
of the IO; its functions, components, and the general layout. 
Section \ref{sec:requirements} discusses the requirements for the IO.  
Section \ref{sec:components} presents the core of this paper; it will describe individual
IO components, their performance in pre-installation tests and the
detailed layout of the IO. Section \ref{sec:performance} discusses the expected
and measured in-vacuum performance as known by the time of writing.
The final integrated testing of the IO subsystem at design sensitivity
requires the main interferometer to be nearly fully commissioned to
act as a reference for many of the required measurements; this will
be discussed in Section \ref{sec:conclusion}.

%==================================================================================================
%  Overview of the Input Optics
%==================================================================================================
\section{Overview of the Input Optics }
\label{sec:overview}

\begin{figure}[t]
	\centering
	\includegraphics[width=\linewidth]{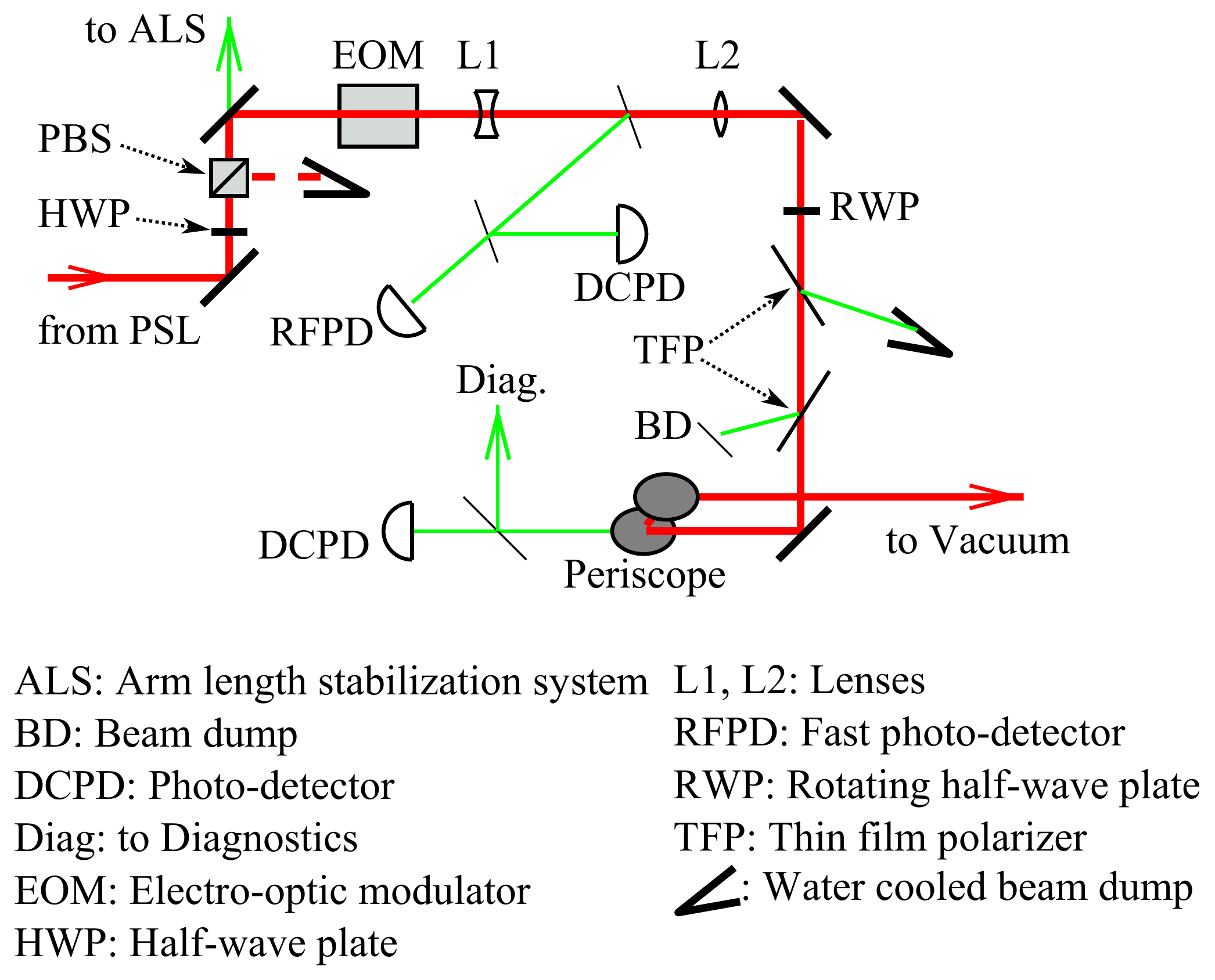}
	\caption{\label{fig:PSL-IO} The IO on the PSL modulates the phase of the laser
		beam with the EOM, mode matches the light into the input mode cleaner
		(located inside the vacuum system), and controls the power injected
		into vacuum system.}
\end{figure}

Figure \ref{fig:PSL-IO} shows a sketch of the first part of the input
optics. This part is co-located with the PSL on the same optical table
inside the laser enclosure, outside of the vacuum system. 
It prepares the laser beam for the injection
into the vacuum system. The beam from the PSL is first routed through
a half wave plate and a polarizing beam splitter. These two elements
form a manual power control stage which is used mainly
during alignment processes on the optical table. The following mirror
transmits 2.5\% of the light. This light is used by the arm length stabilization 
(ALS) system during lock acquisition of the main interferometer
\cite{mullavey_arm-length_2012}. 

Most of the light is sent through an electro-optic
modulator (EOM) which modulates the phase of the laser field with three
different modulation frequencies. Two of these frequencies are used
by the interferometer sensing and control system (ISC) to sense most of
the longitudinal and alignment degrees of freedom of the mirrors inside
the IFO and to stabilize the laser frequency and the alignment of
the laser beam into the interferometer. The third frequency is used
to control the input mode cleaner (IMC). The two lenses L1 and L2 mode match
the beam to the in-vacuum input mode cleaner. The next steering mirror
directs the beam through another half wave plate inside a motorized
rotation stage in front of two thin film polarizers. This second power
control stage is used during operations to adjust the power to
the requested level. The periscope raises the height of the beam and 
steers it into the vacuum system. The top mirror is mounted on a
piezo actuated mirror mount to fine tune the alignment
of the beam into the vacuum chamber.

Between the lenses is a wedge to pick off a small fraction of
the laser beam for diagnostic purposes. A fast photo detector monitors
the residual amplitude modulation at the phase modulation frequencies
while a second photodetector monitors the DC power. A fraction of
the main beam also transmits through the bottom periscope mirror and
is used to monitor the power going into the vacuum system
as well as the size, shape, and quality of the beam. 

%^^^^^^^^^^^^^^^^^^^^^^^^^^^^^^^^^^^^^^^^^^^^^^^^^^^^^^^^^^^^^^^^^^^^^^^^^^^^^^^^^^^^^^^^^^^^^^^^^^
%^^^^^^^^^^^^^^^^^^^^^^^^^^^^^^^^^^^^^^^^^^^^^^^^^^^^^^^^^^^^^^^^^^^^^^^^^^^^^^^^^^^^^^^^^^^^^^^^^^

\begin{figure*}
\begin{centering}
\includegraphics[width=\linewidth]{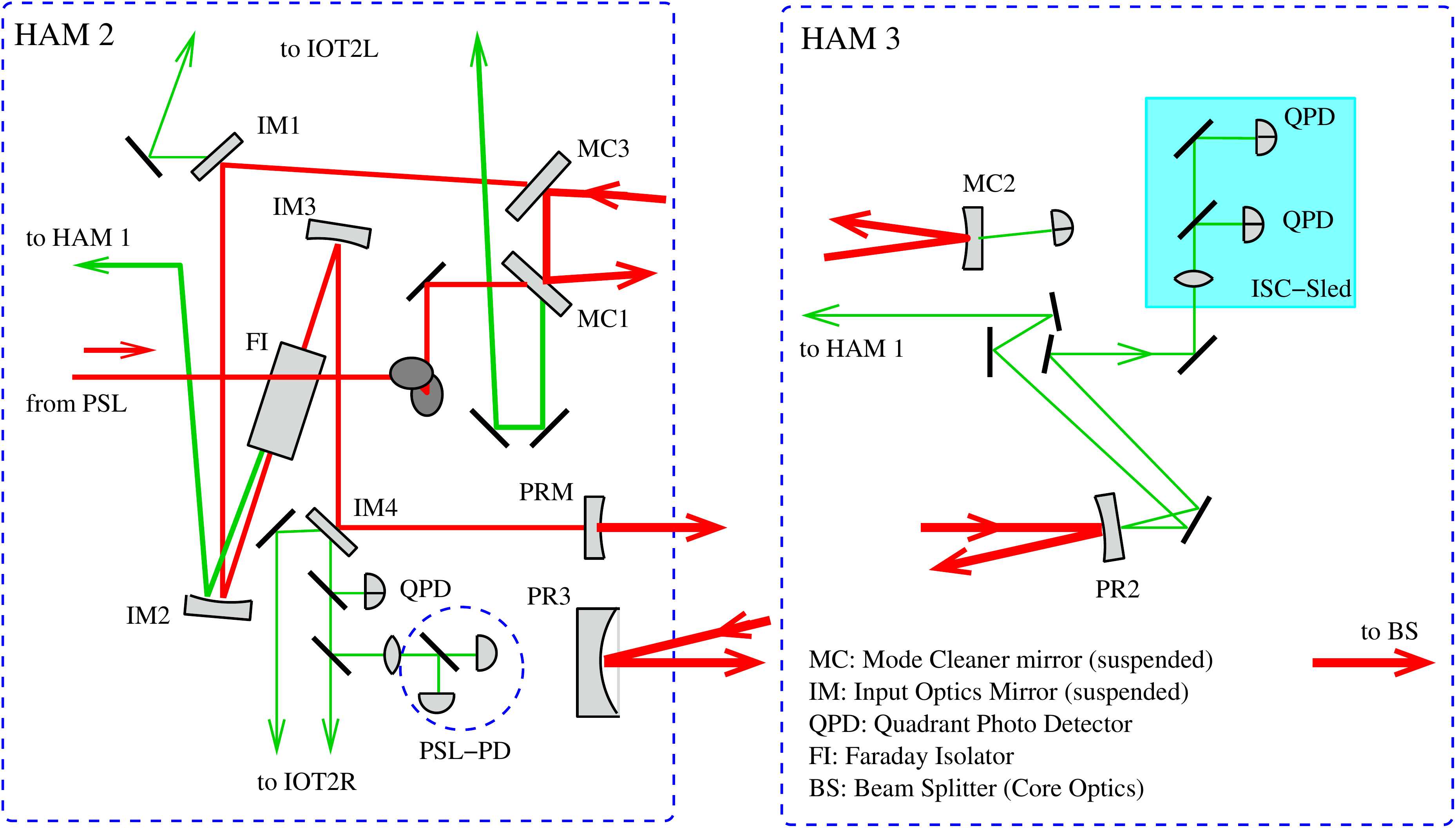}
\par\end{centering}

\protect\caption{\label{fig:The-in-vacuum-parts}A sketch of the in-vacuum components
and beam directions within the input optics in HAM 2 and HAM 3.
The red beam is the forward going main beam while the green beams
are auxiliary beams. The main items in the in-vacuum input optics are the input mode
cleaner (IMC) which is formed by the three mirrors MC1, MC2, MC3;
the Faraday isolator (FI); and the four suspended steering mirrors IM1-4
of which IM2 and 3 match the spatial mode of the IMC into the main
interferometer. The recycling cavity mirrors PRM, PR2, and PR3 are
not part of the input optics. The ISC sled in HAM3 belongs to the
interferometer sensing and control subsystem and provides alignment
signals for the recycling cavity.}

\end{figure*}

Following the periscope, the main beam is sent through a metal tube
which includes a mechanical shutter and through HAM1%
\footnote{HAM: Horizontal Access Modules are nearly spherical vacuum chambers
with a diameter of about 2.8~m.} into HAM2; all in-vacuum IO components are mounted on seismically
isolated optical tables inside HAM2 and HAM3. As shown in Figure
\ref{fig:The-in-vacuum-parts}, the beam passes over the FI
to a second periscope which lowers the beam to the in-vacuum beam height.
The next element in the IO is the suspended IMC, a 33~m long triangular
cavity. 
The two flat input and output mirrors, named MC1 and MC3 respectively, are located in HAM2 while 
the third curved mirror, MC2, is located in HAM3.
Following MC3 are two suspended mirrors, IM1 and IM2,
which steer the beam through the FI. IM3 and IM4
are used to steer the beam into the PRC. IM2 and
IM3 are curved to mode match the output mode of the IMC to the mode of the main
interferometer. 

%^^^^^^^^^^^^^^^^^^^^^^^^^^^^^^^^^^^^^^^^^^^^^^^^^^^^^^^^^^^^^^^^^^^^^^^^^^^^^^^^^^^^^^^^^^^^^^^^^^
%^^^^^^^^^^^^^^^^^^^^^^^^^^^^^^^^^^^^^^^^^^^^^^^^^^^^^^^^^^^^^^^^^^^^^^^^^^^^^^^^^^^^^^^^^^^^^^^^^^

Two of the steering mirrors, IM1 and IM4, transmit a small fraction
of the light creating three different auxiliary beams which are used
to monitor the power and spatial mode of the IMC transmitted beam,
of the beam going into the IFO, and of the beam which is reflected
from the IFO. The latter two beams are routed to IOT2R%
\footnote{IOT2R:~Input Optics Table on the Right side of HAM2}, 
an optical table on the right side of HAM2, while the first beam
and the field which is reflected from MC1 are routed to IOT2L on
the left side of HAM2. The position of the forward going beam through
IM4 is monitored with an in-vacuum quadrant photo detector while
a large fraction of this beam is also sent to an in-vacuum photo detector
array which is used to monitor and stabilize the laser power before
it is injected into the IFO. Most of the IFO reflected
field goes back to the FI where it is separated from
the incoming beam. This field is routed into HAM1 where it is detected
to generate length and alignment sensing signals. 

In HAM3, a small fraction of the intra-mode cleaner field transmits
through MC2 onto a quadrant photo detector to monitor
the beam position on MC2. The forward and backwards traveling waves
inside the PRC partly transmit through PR2 and are routed into HAM1
and to an optical breadboard inside HAM3, respectively. These beams
are used by ISC for sensing and control of the interferometer and
for diagnostic purposes. The breadboard holds two lenses which
image the beam with orthogonal Gouy phases onto two quadrant photo
detectors to monitor beam position and pointing inside the power
recycling cavity. IOT2R and IOT2L host photo detectors and digital
cameras to monitor the power and beam sizes in each of the picked-off
beams. IOT2L also hosts the photo
detectors which are used by the interferometer sensing and control
system to generate length and alignment sensing signals for the input
mode cleaner. 

While the figure shows all key components in the correct sequence,
we intentionally left out the detailed beam routing, the baffles used
to protect all components from the laser beam in case of misalignments,
and the beam dumps to capture all ghost beams. 

A complete document tree which
contains all design and as-built layouts as well as drawings of all
components is available within the LIGO Document Control 
Center\footnote{The LIGO Document Control Center (DCC) is an on line repository of documents 
produced by members of LIGO and the LIGO Scientific Collaboration.  It is accessible from any 
modern web browser at the url https://dcc.ligo.org}
(DCC) under document number E1201013\cite{dcc_e1201013}.

%==================================================================================================
%  Input Optics Requirements
%==================================================================================================
\section{Input Optics requirements }
\label{sec:requirements}

The aLIGO interferometer can be operated in different modes
to optimize the sensitivity for different sources\cite{dcc_t1000298}.
These modes are characterized by the input power and the microscopic position and
reflectivity of the signal recycling mirror. The requirements for
the aLIGO input optics are specified to simultaneously meet
the requirements for all anticipated science modes and address all
degrees of freedom of the laser field. Requirements in aLIGO
are defined for three distinct frequency ranges: DC, the control band
up to 10~Hz, and the signal or detection band from 10~Hz to a few
kHz. The requirements in the detection band are defined in terms of
linear spectral densities and include a safety factor of ten for all
technical noise sources. To first order, a perfectly symmetric Michelson
interferometer is insensitive to all input noise sources which is an often
overlooked reason for its use in the first place. However, all degrees
of freedom of the injected laser field couple via some asymmetry to
the output signal. This drives the requirements in the control band
which are usually defined as RMS values. The more critical requirements
for the IO are:

%************************************************
% Paragraph: Power
%************************************************
\paragraph*{Power:}

The high power science modes require to inject 125~W of mode matched
light into the interferometer with less than an additional 5\% in
higher order modes. The PSL has to deliver 165~W of light in an appropriate
$\text{TEM}_{00}$ mode. Consequently, the net efficiency of $\mbox{TEM}_{00}$
optical power transmission from the PSL output to the main interferometer
has to be above 75\%. This sets limits on accumulated losses in all
optical components but also limits the allowed thermal lensing in
the EOM, the FI, and the power control stages; the reflective
optics and fused silica lenses are much less susceptible to thermal
lensing. Efficient power coupling is also dependent on good mode matching 
between the recycling cavities and the arm cavities in the main interferometer.

%************************************************
% Paragraph: Power Control
%************************************************
\paragraph*{Power Control:}

The injected power into the interferometer has to be adjustable from
the control room from minimum to full power for diagnostic and operational
purposes, to acquire lock of the main interferometer, and to operate
between different science modes. The rate of power change (dP/dt)
has to be sufficiently small to limit the radiation pressure kick
inside the IMC and the main interferometer to a level that can be
handled by the length and alignment control system. It has to be sufficiently
fast to not limit the time to transition to full power after lock acquisition; 
i.e. it should be possible to change from minimum to maximum power within a few seconds. 

Note that minimum power here can not mean zero power because of the
limited extinction ratio of polarizers. Going to zero power requires
actuation of the aforementioned mechanical shutter which can only be accessed manually
between the laser enclosure and HAM1. The emergency shutter is part
of the PSL laser system and cuts the laser power at the source. Furthermore,
the power control system within the IO is not used for actively stabilizing
the laser power within the control or the detection band.

%************************************************
% Paragraph: Power Fluctuations
%************************************************
\paragraph*{Power fluctuations:}

Fluctuations in the laser power can couple through many different
channels to the error signal. The noise scales with the asymmetries
in the interferometer. Two different mechanisms are expected to dominate
the susceptibility of the interferometer to power fluctuations. The
optical power inside the arm cavities will push the test masses outwards.
Any change in power will cause fluctuations in that pressure which
can lead to displacement noise at low frequencies. The susceptibility
to radiation pressure noise scales with differences in the power build
up inside the arm cavities and it is assumed that these differences
are below 1\%. At high frequencies, the signal itself limits the allowed
power fluctuations. In lock the two arm cavities are detuned by a few
pm which causes some light to leak out to the dark port\cite{fricke_dc_2012}. 
Gravitational waves will modulate these offsets causing
the light power to fluctuate. Obviously, power fluctuations in the
laser itself, although highly filtered by the interferometer, will
cause similar fluctuations. The relative intensity noise in the detection
band has to be below $2\times10^{-9}/\sqrt{\mbox{Hz}}$ at 10~Hz
increasing with $f$ to $2\times10^{-8}/\sqrt{\mbox{Hz}}$ at 1~kHz
and remaining flat after this\cite{dcc_t0900630}. Furthermore, the expected seismically
excited motion of the test masses limits the allowed radiation pressure
noise in the control band to $10^{-2}/\sqrt{\text{Hz}}$ below 0.2~Hz.
Above 0.2~Hz, the requirements follow two power laws; initially $f^{-7}$
then $f^{-3}$, before connecting with the detection band requirement
at 10~Hz. 

The IO does not provide any active element to change or stabilize
the laser power within the control or the detection band. The PSL
uses a first loop which stabilizes the laser power measured
with a photo-detector on the PSL table to $2\times10^{-8}/\sqrt{\mbox{Hz}}$
between 20 and 100~Hz and meeting the aforementioned requirements
above 100~Hz. The PSL also measures the power of the injected field
with the photo detector array shown (PSL-PD) in Figure \ref{fig:The-in-vacuum-parts}.
The IO has to supply this auxiliary beam and maintain a sufficiently
high correlation with the injected beam and minimize the chances of
additional power fluctuations within any of these two beams.

%************************************************
% Paragraph: Frequency Fluctuations
%************************************************
\paragraph*{Frequency fluctuations:}

In the detection band, the laser frequency will ultimately be stabilized
to the common mode of the two arm cavities which are the most stable
references available in this frequency range. At lower frequencies
the arm cavities are not a good reference and are made to follow the frequency
reference inside of the PSL. The input mode cleaner acts as a frequency
reference during lock acquisition and as an intermediate frequency
reference during science mode. It is integrated into the complex and
nested laser frequency stabilization system. Based on the expected
common mode servo gain the frequency noise requirements for the IMC
are set to:

\begin{align*}
	\delta\nu(f=10\,\mbox{Hz})&<50\frac{\mbox{mHz}}{\sqrt{\mbox{Hz}}}\\
	\delta\nu(f\geq100\,\mbox{Hz})&<1\frac{\mbox{mHz}}{\sqrt{\mbox{Hz}}}.
\end{align*}

These requirements can be expressed equivalently as length fluctuations of 
the IMC:
\begin{align*}
	\delta\ell(f=10\,\mbox{Hz})&<3\cdot10^{-15}\frac{\mbox{m}}{\sqrt{\mbox{Hz}}}\\
	\delta\ell(f\geq100\,\mbox{Hz})&<6\cdot10^{-17}\frac{\mbox{m}}{\sqrt{\mbox{Hz}}}.
\end{align*}

%************************************************
% Paragraph: Frequency
%************************************************
\paragraph*{RF modulation frequencies:}

The main laser field consists of a carrier and multiple pairs of sidebands.
The carrier has to be resonant in the arm cavities and the power recycling
cavity; the resonance condition in the signal recycling cavity depends
on the tuning and specific science mode. One pair of sidebands must 
be resonant in the power recycling cavity while the second pair must resonate in
both the power and signal recycling cavity.
The modulation signals of $f_{1}=9.1\,\mbox{MHz}$
and $f_{2}=45.5\,\mbox{MHz}$ are provided by the interferometer sensing
and control system. A third modulation frequency of $f_{3}=24.1\,\mbox{MHz}$
is required to sense and control the input mode cleaner. The last
pair of sidebands should be rejected by the input mode
cleaner so as not to interfere with the sensing and control system of the main interferometer.

%************************************************
% Paragraph: RF Modulation
%************************************************
\paragraph*{RF modulation depth:}

The required modulation depths depend on the final length and alignment sensing
and control scheme. This scheme is likely to evolve over the commissioning
time but the current assumption is that a modulation index of 0.4
for a 24dBm signal driving the EOM is more than sufficient. Note that
this only applies to the two modulation frequencies which are used
for sensing and control of the main interferometer; the modulation
index for the third frequency need only be large enough to control the
IMC. 

The classic phase modulation/demodulation sensing scheme for a single
optical cavity measures how much the cavity converts phase modulation
into amplitude modulation when near resonance. Unfortunately all phase modulators also
modulate the amplitude of the laser field. This amplitude modulation
can saturate the RF amplifiers and mixers in the detection chain and
generate offsets in the error signals which have to be compensated.
aLIGO requires that the amplitude modulation index is less
than $10^{-4}$ of the phase modulation index.\cite{kokeyama_residual_2014}

%************************************************
% Paragraph: RF Modulation Noise
%************************************************
\paragraph*{RF modulation noise:}

Changes in the amplitude and phase of the RF modulation signals can
change the dark port signal by changing the power build-up of the
carrier in the arm cavities or through cross coupling in the length
and alignment sensing and control schemes. These effects were analyzed by the ISC 
group\cite{dcc_t1000298}. The analysis uses specifications from a commercial crystal
oscillator manufacturer as the expected oscillator phase and amplitude
noise. These specifications for phase noise are $10^{-5}\mbox{rad}/\sqrt{\mbox{Hz}}$
at 10~Hz falling with $1/f^{3/2}$ to $3\times10^{-7}\mbox{rad}/\sqrt{\mbox{Hz}}$
at 100~Hz and then a little faster than $1/f$ to $2\times10^{-8}\mbox{rad}/\sqrt{\mbox{Hz}}$
at a kHz above which the requirement stays constant. The specifications
for amplitude noise are $10^{-7}/\sqrt{\text{Hz}}$ at 10~Hz falling
with 1/f between 10 and 100~Hz and then with $1/\sqrt{f}$ until
1~kHz above which it stays constant at $3\times10^{-9}/\sqrt{\mbox{Hz}}$.
These specifications have been adopted as requirements although the
analysis shows that they could be relaxed at higher frequencies.

%************************************************
% Paragraph: Beam Jitter
%************************************************
\paragraph*{Beam jitter:}

Changes in the location and direction of the injected beam can be
described as scattering light from the $\mbox{TEM}_{00}$ into a $\mbox{TEM}_{10}$
mode. This light scatters back into the $\mbox{TEM}_{00}$ mode inside
a misaligned interferometer and creates noise in the gravitational
wave signal.\cite{mueller_beam_2005} This is an example where noise in
the detection band, here beam jitter, couples to noise in the control
band, here tilt of the input test masses. It is expected that the
test masses will all be aligned to better than 2~nrad RMS with respect
to the nominal optical axis of the interferometer. Under this assumption,
the relative amplitude of the injected 10-mode has to stay below $10^{-6}/\sqrt{\mbox{Hz}}$
at 10~Hz falling with $1/f^{2}$ until 100~Hz above which the requirement
stays constant at $10^{-8}/\sqrt{\mbox{Hz}}$.

%************************************************
% Paragraph: Optical Isolation
%************************************************
\paragraph*{Optical isolation:}

The FI isolates the IMC from back reflected light from
the main interferometer. The requirements for the isolation ratio
are based on experience gained during the initial years of operating
LIGO and also VIRGO. Virgo operated for a long time without a FI between the mode cleaner and the main interferometer and
encountered problems due to the uncontrolled length between the IMC and 
IFO\cite{acernese_results_2004} (a parasitic interferometer). Initial
and enhanced LIGO never encountered any major problems with insufficient
optical isolation in the FI. The requirements of 30dB
for the optical isolation in the FI were set based on
the experience in initial LIGO, taking into account the higher injected
power.

%************************************************
% Paragraph: Additional Requirements
%************************************************
\paragraph*{Additional requirements:}

It is well known that parasitic interferometers and scattered light
together with mechanically excited surfaces can add frequency and
amplitude noise to a laser beam. The IO adopted a policy to limit
the added noise to 10\% of the allowed noise; note that the allowed
frequency and amplitude noise prior to the input mode cleaner is significantly
higher than after the mode cleaner. This drives requirements on
the residual motion of the optical components, the surface quality
of all optical components and their coatings, and on the placement
and efficiency of the optical baffles. The requirement to align the
IO drives requirements on actuation ranges for all optics and, last
but not least, the IO has to meet the stringent cleanliness and vacuum
requirements of aLIGO. These requirements are discussed throughout the 
paper when relevant. 

The official requirements were originally captured in LIGO document
T020020\cite{dcc_t020020} and have been derived, updated and clarified in several other
documents which are all available under E1201013.\cite{dcc_e1201013}

%==================================================================================================
%  Input Optics Components and Final Layout
%==================================================================================================
\section{Input Optics components and final layout}
\label{sec:components}

This section will first discuss the individual components and their
measured performance. This will be followed by a description of the
optical layout which includes a discussion of beam parameters and
mode matching between the various areas. 

%--------------------------------------------------------------------------------------------------
% Electro-optic Modulators
%--------------------------------------------------------------------------------------------------
\subsection{Electro-optic modulators}

The EOM must use a material capable of withstanding CW optical powers
of up to 200~W and intensities up to $25\,\mbox{kW}/\mbox{cm}^{2}$.
At these power levels the induced thermal lensing, stress induced
depolarization, and damage threshold of the electro-optic material
must be taken into consideration. Rubidium titanyl phosphate (RTP)
was chosen many years ago over other electro-optic materials, such as
rubidium titanyl arsenate (RbTiOAsO4 or RTA) and lithium niobate (LiNb03), 
as the most promising
modulator material after a literature survey, discussions with various
vendors, and corroborating lab experiments.\cite{dcc_t060267, albrecht_study_2006} RTP has
a very high damage threshold, low optical absorption, and a fairly high
electro-optical coefficient. Enhanced LIGO (eLIGO) allowed for testing of the material
and design over a one-year period at 30~W input power.\cite{dooley_thermal_2012} 

%^^^^^^^^^^^^^^^^^^^^^^^^^^^^^^^^^^^^^^^^^^^^^^^^^^^^^^^^^^^^^^^^^^^^^^^^^^^^^^^^^^^^^^^^^^^^^^^^^^
\begin{figure}[t]
\begin{centering}
\includegraphics[height=4.2cm]{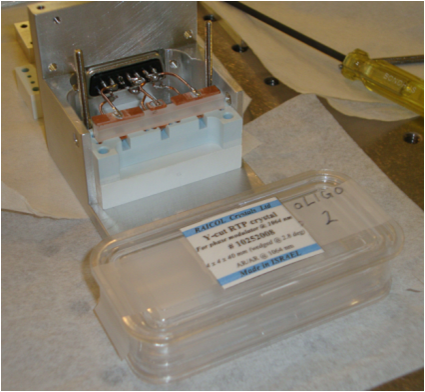}$\quad$\includegraphics[height=4.2cm]{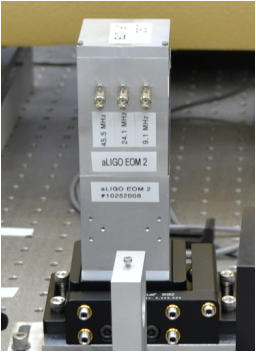}
\par\end{centering}

\protect\caption{\label{fig:aLIGO-EOM1}
	Images of the EOM.  
	The housing uses two modules; 
	the crystal and the electrodes are placed in the lower module while the upper module houses the 
	coils for the three resonant circuits. 
	The left picture shows the inside of the lower module: the aLIGO EOM consists of a 
	wedged RTP crystal with three pairs of electrodes. 
	The two 15~mm electrodes on the outside are used for the main modulation frequencies 
	$f_{1}=9.1\,\mbox{MHz}$ and $f_{2}=45.5\,\mbox{MHz}$. 
	The 7~mm electrodes in the middle are used for $f_{3}=24.1\,\mbox{MHz}$. 
	The crystal and the electrodes are clamped between two macor pieces. 
	The right picture shows the final modulator (both modules) on a five axis alignment stage. 
	}
\end{figure}
%^^^^^^^^^^^^^^^^^^^^^^^^^^^^^^^^^^^^^^^^^^^^^^^^^^^^^^^^^^^^^^^^^^^^^^^^^^^^^^^^^^^^^^^^^^^^^^^^^^

The aLIGO EOM uses a patented design\cite{quetschke_method_2013} which is very similar to the 
one used in eLIGO; both 
consist of a $4\times4\times40\,\mbox{mm}$ long wedged RTP
crystal (see Figure \ref{fig:aLIGO-EOM1}). The $2.85^{\circ}$ wedges
prohibit parasitic interferometers from building up inside the crystal and
allow for separation of the two polarizations of the injected laser field
with an extinction ratio of better than $10^{5}$. This separation
avoids polarization rotation which could otherwise convert phase modulation
to amplitude modulation. The AR coated surfaces have a reflectivity
of less than 0.1\%. For aLIGO we use two 15~mm long pairs
of electrodes for the two main modulation frequencies and one 7~mm
long pair for the auxiliary frequency used to control the IMC. Each
electrode pair forms a capacitor which is part of a resonant circuit
in the form of a $\pi$ network where the additional inductor and
capacitor are used to simultaneously match the resonance frequency
and create the required $50\,\Omega$ input impedance.

%^^^^^^^^^^^^^^^^^^^^^^^^^^^^^^^^^^^^^^^^^^^^^^^^^^^^^^^^^^^^^^^^^^^^^^^^^^^^^^^^^^^^^^^^^^^^^^^^^^
\begin{figure}[t]
	\centering
	\includegraphics[width=\linewidth]{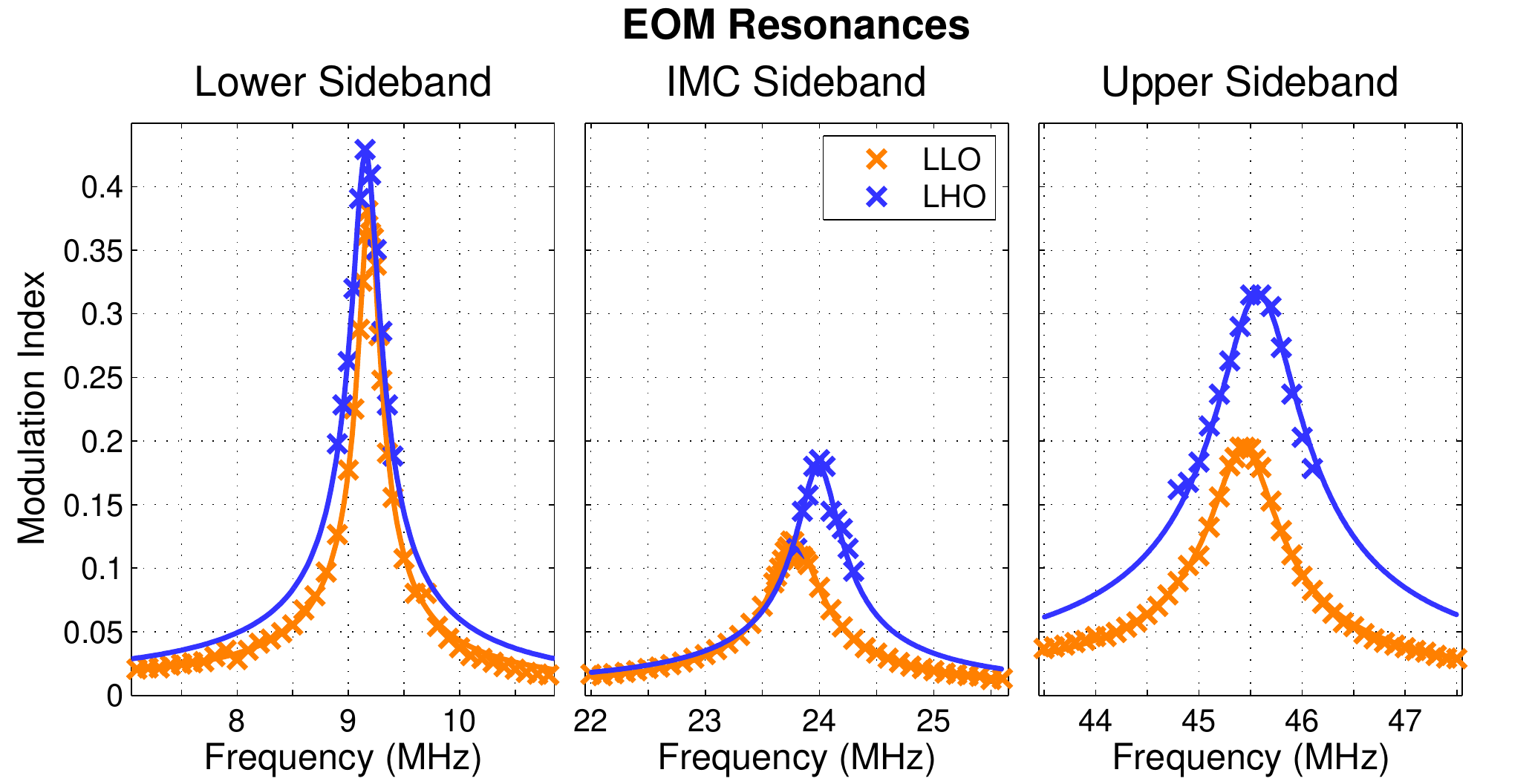}
	\caption{\label{fig:aLIGO-EOM2}
		The measured modulation indices for both the Livingston and Hanford EOM with a 24 dBm drive.  
		The data is shown together with a best fit to the expected circuit response.}
\end{figure}
%^^^^^^^^^^^^^^^^^^^^^^^^^^^^^^^^^^^^^^^^^^^^^^^^^^^^^^^^^^^^^^^^^^^^^^^^^^^^^^^^^^^^^^^^^^^^^^^^^^
	
After installation and alignment at both sites, initial tests confirmed
that the RTP crystals do not produce a significant thermal lens.
An optical spectrum analyzer was used to measure the modulation index 
as a function of modulation
frequency for each of the three resonant circuits. The results are
shown in Figure \ref{fig:aLIGO-EOM2}. The modulation indices for $f_1$ and 
$f_2$ meet the requirements at both sites while the $f_3$ modulation index is still a little 
low, especially at LLO. However,
early commissioning experience indicates that the modulation indices are sufficient for the
aLIGO length and alignment sensing scheme and it was decided
to use the EOM as is for now and potentially improve the resonant
circuits later if necessary. 

The residual amplitude modulation (RFAM) produced by the EOM was also characterized.  
The AM/PM ratio for each of the three sidebands was measured to be 
$1.0\cdot10^{-4}$, $1.2\cdot10^{-5}$, and $4.1\cdot10^{-5}$ for the 9.1 MHz, 24.0 MHz and 
45.5 MHz sidebands respectively.  
All three measurements come out to be at or below the requirement of $10^{-4}$ derived by 
Kokeyama et. al.\cite{kokeyama_residual_2014}  
The temperature dependence of the RFAM generation was briefly investigated and was found to 
be able to push the AM/PM ratio at 9.1 MHz up as high as $3\cdot10^{-4}$.  
This may need to be addressed with a temperature stabilization servo in the future if RFAM is 
found to be an issue during detector commissioning, but the design of the modulator was left 
unchanged until such an issue arises.  

Detailed design drawings, assembly instructions,
and test reports are available under LIGO document number T1300084.\cite{dcc_t1300084}

%--------------------------------------------------------------------------------------------------
% Faraday Isolator
%--------------------------------------------------------------------------------------------------
\subsection{Faraday isolator}

The FI is a much more complicated optical device compared
to the EOM. It is more susceptible to thermal lensing and its location
after the mode cleaner amplifies the requirement to maintain a good
spatial mode. The FI has to handle between 20 to 130~W
of laser power without significantly altering the beam profile or polarization of
the beam. 
Like the EOM, the aLIGO FI is also very similar to the FI used
in eLIGO\cite{dooley_thermal_2012}. Both were designed to minimize
and mitigate thermal lensing and thermal stress induced depolarization
by compensating these effects in 
subsequent crystals\cite{snetkov_compensation_2011,khazanov_compensation_2004}. 

%^^^^^^^^^^^^^^^^^^^^^^^^^^^^^^^^^^^^^^^^^^^^^^^^^^^^^^^^^^^^^^^^^^^^^^^^^^^^^^^^^^^^^^^^^^^^^^^^^^
\begin{figure}[t]
\begin{centering}
\includegraphics[clip,width=\linewidth]{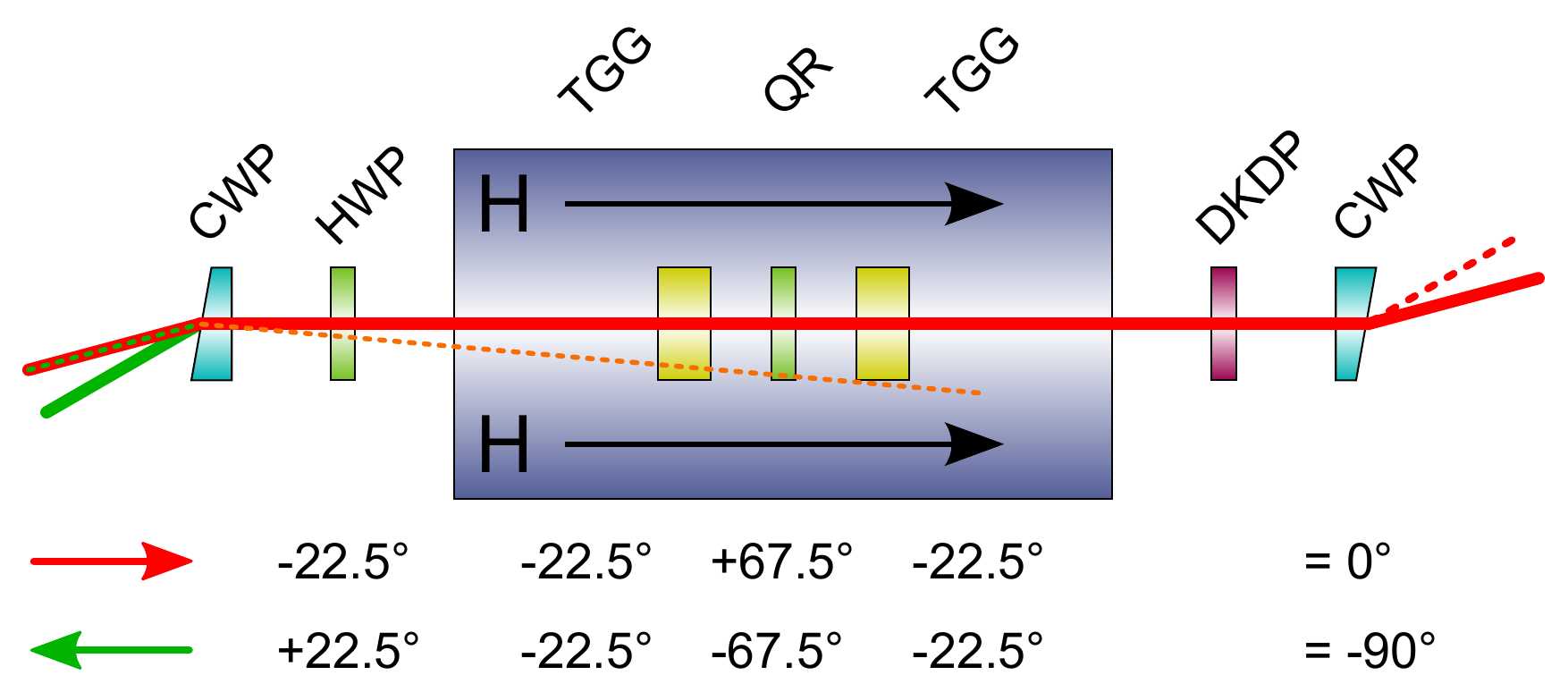}
\par\end{centering}

\begin{centering}
\includegraphics[clip,width=\linewidth]{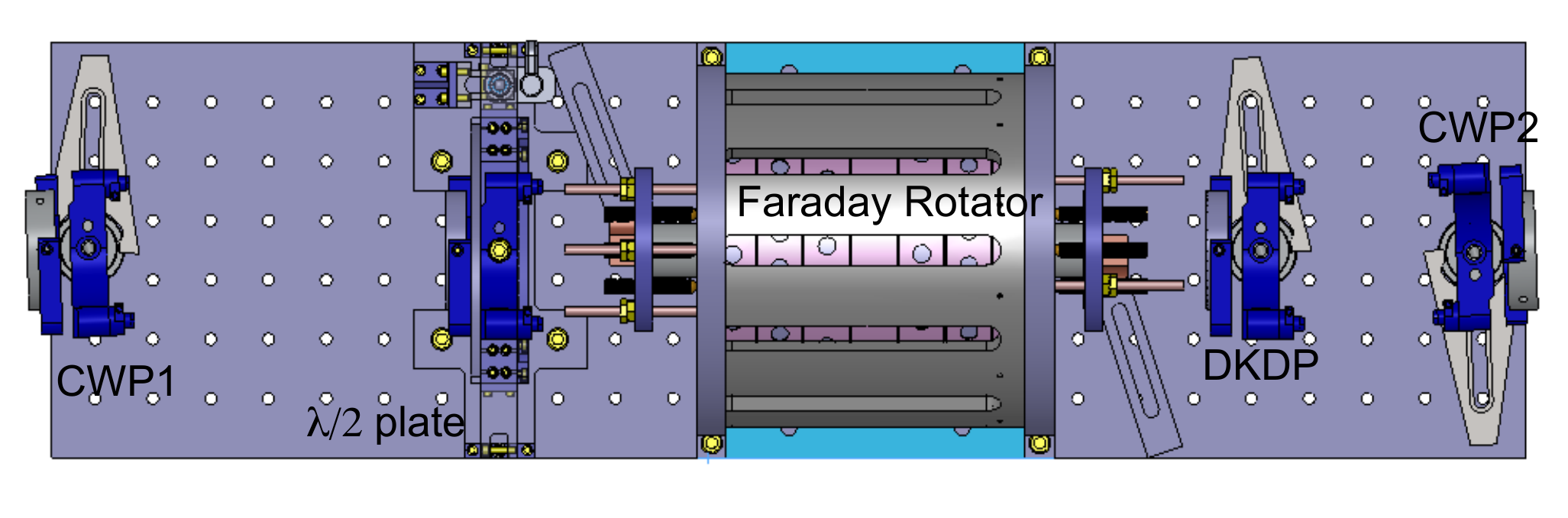}
\par\end{centering}

\begin{centering}
\includegraphics[clip,width=\linewidth]{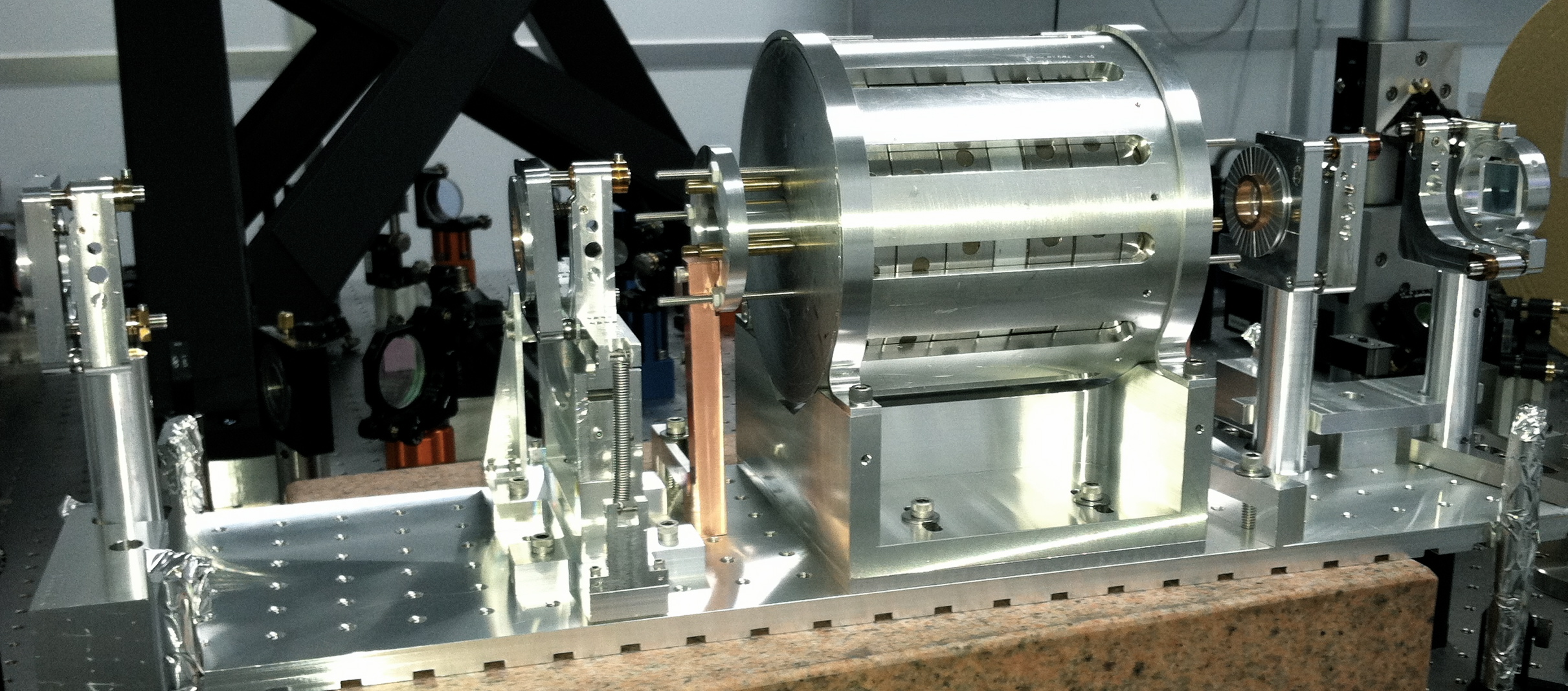}
\par\end{centering}

\protect\caption{\label{fig:Advanced-LIGO-Faraday}Advanced LIGO Faraday isolator (from
top to bottom): optical layout, design, final product.}

\end{figure}
%^^^^^^^^^^^^^^^^^^^^^^^^^^^^^^^^^^^^^^^^^^^^^^^^^^^^^^^^^^^^^^^^^^^^^^^^^^^^^^^^^^^^^^^^^^^^^^^^^^

The aLIGO FI design consists of a Faraday rotator, a pair
of calcite-wedge polarizers, an element with a negative dn/dT for
thermal-lens compensation, and a picomotor-controlled half-wave plate
for restoring the optical isolation in-situ. In addition, a heat sink
is connected to the holders of the magneto-optical crystals to drain 
excess heat into the FI breadboard. 
The Faraday rotator is based on an arrangement developed
by Khazanov et. al.\cite{khazanov_suppression_2000}, that uses a pair of 
\tildegood1~cm
long Terbium Gallium Garnet (TGG) crystals as magneto-optical elements,
each nominally producing a $22.5^{\circ}$ rotation of the electric
field when placed in a magnetic field of about 1T. They are separated
by a \tildegood1~cm 
 long piece of quartz that rotates the polarization
field reciprocally by $67.5^{\circ}\pm0.6^{\circ}$ . This arrangement
(shown schematically at the top of Figure \ref{fig:Advanced-LIGO-Faraday}) allows
thermally induced birefringence produced in the first magneto-optical
element to be mostly compensated in the second one. The HWP is a zero-order
epoxy-free quartz half-wave plate. It is set to rotate the polarization
by an additional $22.5^{\circ}$ to have $0^\circ$ net rotation in the
forward going and $90^{\circ}$ in the backward going direction.

All crystals were selected to minimize absorption, thermal beam distortion
and surface roughness. 
Those made of harder and non-hygroscopic materials; the half-wave plate,
quartz rotator, and TGG crystals, are all super-polished (surface roughness
below 0.5~nm) and received a custom low loss IBS AR coating with a
rest reflectivity of less than 300~ppm. The softer calcite polarizers and the DKDP
crystal were procured from the manufacturers with their standard polishings
and coatings. The two calcite polarizers each have a thickness of
\tildegood5~mm and are wedged at $8.5^{\circ}$ to allow the
orthogonally polarized beams to separate sufficiently. The calcite
wedges have an extinction ratio of at least $10^{5}$ and more than
99\% optical efficiency. 

The magnetic field is created by a stack of seven magnetized Fe-Nb
magnetic disks\cite{shiraishi_compact_1986} each having a bore of 24~mm and a thickness of
19.7~mm. This stack produces a maximum axial field of 1.16~T
(LLO) and 1.55~T (LHO) near its center which falls off towards the
end. The difference in the magnetic field is caused by the selection
of the magnetic materials and the thermal treatment of the individual
magnets.\cite{palashov_high-vacuum-compatible_2012} The TGG crystals and
quartz rotator are installed about 3~cm apart from each other before 
being fine tuned to produce $22.5^\circ$ of rotation by adjusting their depth in the magnet. The
entire FI is mounted on a 648 mm x 178 mm breadboard for convenient
transfer into the HAM chamber after out-of-vacuum optimization.

After undergoing a thorough cleaning procedure, 
the FI was assembled and aligned with the main PSL beam 
in the laser enclosure. The optical table in the
enclosure is made from stainless steel while the optical table in
HAM2 is made from aluminum. The differences in magnetic susceptibility
are significant enough to require the FI to be raised with an \tildegood11~cm
thick granite block visible in the bottom picture in Figure \ref{fig:Advanced-LIGO-Faraday}. 
The bottom periscope mirror in Figure \ref{fig:PSL-IO}
was removed and the beam was sent via several mirrors through the
FI. This setting ensured that the beam parameters, beam
size and divergence angle, are very similar to the ones expected
in-vacuum. 

%^^^^^^^^^^^^^^^^^^^^^^^^^^^^^^^^^^^^^^^^^^^^^^^^^^^^^^^^^^^^^^^^^^^^^^^^^^^^^^^^^^^^^^^^^^^^^^^^^^
\begin{figure*}[t]
\begin{centering}
\includegraphics[width=0.48\linewidth]{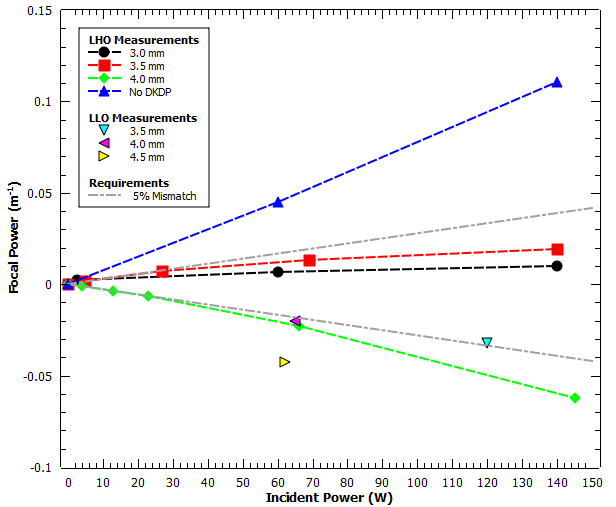}$\qquad$\includegraphics[width=0.48\linewidth]{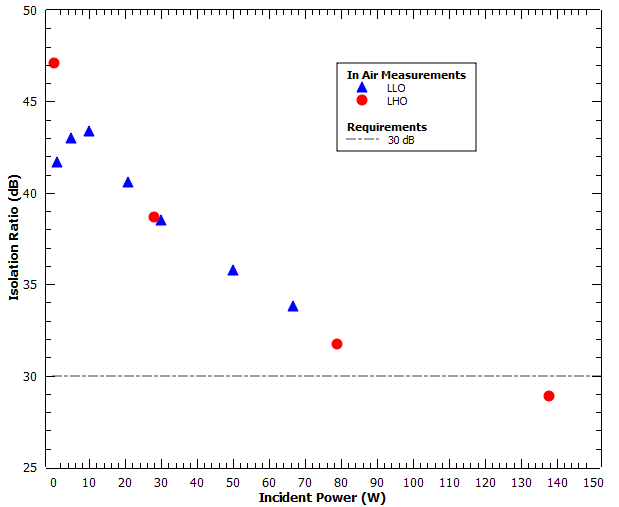}
\par\end{centering}

\protect\caption{\label{fig:Isolation-ratio-measured}Left graph: Thermal lens for
various DKDP crystals measured in-air as a function of laser power. Right
graph: Isolation ratio measured in-air at both sites. The power is
the injected power while the power inside the FI is
twice as high. Therefore 70~W incident power corresponds to \tildegood 125~W
injected power during science mode when the near impedance matched
interferometer reflects less than 10\% of the light.}

\end{figure*}
%^^^^^^^^^^^^^^^^^^^^^^^^^^^^^^^^^^^^^^^^^^^^^^^^^^^^^^^^^^^^^^^^^^^^^^^^^^^^^^^^^^^^^^^^^^^^^^^^^^

The thermal lensing of the FI was determined from
beam-scan measurements of a sample of the beam after it was transmitted
through the isolator for incident powers as high as 120 W at LLO and
140 W at LHO. At both sites, the diagnostic beam was focused with
a lens of 1 m focal length and the beam profile was recorded with
CCD or rotating slit beam scans as a function of power for different DKDP crystals. 
The thermal lens at the location of the FI was then computed using
an ABCD matrix algorithm. Figure \ref{fig:Isolation-ratio-measured} shows
the thermal lensing measurements for the TGG crystals and different
DKDP crystals at LHO and LLO. The length of the DKDP crystal
was chosen to compensate the a priori unknown thermal lensing in the
TGG crystals. Based on experience from initial and enhanced LIGO,
the expectation was that DKDP crystals between $3.5\,\text{mm}$ and
$5.5\,\mbox{mm}$ would be needed to compensate the thermal lensing
in the TGG crystals. However, the absorption in the newly purchased TGG
crystals was lower than expected and even our shortest crystals overcompensated.
While the low absorption in TGG is obviously good, it required to
shorten the originally ordered DKDP. We choose 3.5~mm for both isolators
instead of the more optimum 3~mm because of concerns that a thinner
DKDP crystal might fracture inside the vacuum chamber under thermal
stress. Both isolators meet the thermal lensing requirements for aLIGO.

The isolation ratio was also measured as a function of input power.  
To do so the transmitted beam was reflected back under a negligibly small angle
to allow to separate the return beam from the incoming beam. The powers
in the beam going into and through the FI, and in the return beams in both polarizations
were measured to determine and optimize the optical efficiency and
the isolation ratio as a function of laser power. The results for
both FI are shown in Figure \ref{fig:Isolation-ratio-measured}.
For these measurements, the power inside the FI is twice the incident power.
70~W incident power or \tildegood140~W inside the FI is more
than the maximum power we expect during science mode when 125~W are
submitted to the near impedance matched main interferometer and less
than 10~W are reflected. These results show that the in-air tested
FI meet the aLIGO requirements. 

One problem with this approach of optimizing the isolation ratio before 
installation into the vacuum system is that the temperature dependence of 
the Verdet constant will cause the rotation of the two TGG crystals to shift 
after installation.  This effect also causes a power dependent 
shift in the rotation angle.  The motorized rotation
stage allows to adjust the half-wave plate to compensate for these
changes and to optimize the isolation ratio in-situ for the injected
power. This in-situ optimization is important because it is planned to 
operate aLIGO with \tildegood20~W of injected power during the first science 
runs.  

Historically, the FI has always been one of the main
sources of optical losses. The aLIGO FI consists of seven
optical elements with a total of fourteen optical surfaces each contributing
to the losses. The TGG crystals and the DKDP crystals are also known
for absorbing a non-negligible fraction of the light, hence the thermal lensing
(which also reduces the power in the fundamental mode). The next culprit
is the polarization rotation between the two polarizers. Ideally,
the FI would have $0^{\circ}$ in the forward and $90^{\circ}$ rotation in
the backward direction. This is only possible when the TGG crystals
provide exactly $45^{\circ}$in both directions and the quartz rotator
and the half-wave plate combined give exactly $\mp45^{\circ}$ in the forward
and backward direction, respectively. Only the isolation ratio or
the optical efficiency can be optimized in-situ by rotating the half-wave
plate to compensate the aforementioned changes in the Verdet constant
when moving from low to high power and from air to vacuum. The measured
optical efficiency of the FI was 96.7\% ($\pm$0.4\%) for up to 70~W
input power at LLO and 97.7\% ($\pm$0.4\%) for up to 140~W at LHO. 

Prior to installation of the LLO FI, the half-wave plate was temporarily 
adjusted to maximize the optical efficiency rather than the isolation ratio. 
By measuring the residual power dumped in transmission of the FI an upper limit was placed on the 
homogeneity of the polarization rotation at 36~ppm.  
Measuring the isolation ratio in this configuration also allows for a measurement of the missing 
rotation in the Faraday rotator which came out to be \tildegood1.6$^\circ$ at LLO.

%--------------------------------------------------------------------------------------------------
% Input Mode Cleaner
%--------------------------------------------------------------------------------------------------
\subsection{Input Mode Cleaner}

The input mode cleaner (IMC) is a resonant triangular cavity consisting of the three mirrors 
MC1, MC2, and MC3 which form an isosceles triangle as shown in Figure \ref{fig:The-in-vacuum-parts}.
The purpose of the IMC within the input optics is multifaceted.  
It suppresses spatial non-uniformities of the input laser beam while transmitting the diffraction 
limited Gaussian mode.  
It passively suppresses frequency and pointing noise and serves as a reference for additional 
active suppression.  
In addition, the IMC filters the polarization of the input beam before being sent to the FI.  
The input and output coupler (MC1 and MC3 respectively) of the IMC are nominally flat with a 
transmissivity of 6000 ppm while the apex mirror (MC2) has a nominal radius of curvature of 27.27 m 
and a transmissivity of 5 ppm.  
MC1 and MC3 are separated by 46.5 cm while the distance between MC2 and MC1/MC3 is 16.24 m.  
This gives the IMC a free spectral range of 9.099 MHz, a finesse of 515, and a cavity pole of 8.72 
kHz.  

The reflected beam from the IMC is brought out-of-vacuum to the IOT2L table where it is detected  
with a narrowband photodetector for length sensing via the Pound-Drever-Hall\cite{drever_laser_1983} (PDH) 
technique.  
Some of the reflected beam is picked off and sent to two wavefront 
sensors\cite{anderson_alignment_1984, fritschel_alignment_1998}, separated 
by $90^\circ$ of Gouy phase, for angular sensing of the IMC.  
The light leaking through MC2 is sent to an in-vacuum quadrant detector for additional 
angular information.  
In addition, a sample of the transmitted light of the IMC is brought out-of-vacuum to the IOT2L 
table for diagnostics.  

The three mirrors of the IMC are made of fused silica and have a mass of 2.9 kg.  
They hang from the aLIGO small triple suspensions\cite{dcc_t0900435} which provide isolation 
from seismic noise proportional to $f^{-6}$ above the three resonant frequencies near 
1~Hz for all degrees of freedom except vertical and roll.  
The vertical and roll degrees of freedom are isolated with blade springs which provide isolation 
proportional to $f^{-4}$ above the two blade spring resonances near 1 Hz.  
Each stage of the suspensions, including the mirror, have small permanent magnets attached which 
can be actuated upon with electromagnets attached to the suspension frame, known as 
OSEMs.\cite{dcc_t0900286,carbone_sensors_2012}  
The OSEMs also incorporate shadow sensors which use the magnets as flags to sense the 
important degrees of freedom of each stage.    
The actuation strength gets progressively stronger at higher stages with the middle and upper stages 
having, respectively, an actuation authority at DC that is \tildegood20 times and \tildegood1500 times 
that of the mirror.    
Staging the actuation strength in this way prevents the applied force from spoiling the seismic 
isolation provided by the suspension.  

Length sensing of the IMC is accomplished with the PDH technique by 
adding a 24.0 MHz sideband to the beam via the EOM and sensing the amplitude 
modulation induced when the cavity is off resonance.  
This signal provides an accurate comparison between the round trip length of the IMC and the 
frequency of the laser which is used to quiet the laser frequency above \tildegood15~Hz and to quiet 
the cavity length below \tildegood15~Hz.  
Controlling the cavity length employs hierarchical control in which control at lower frequencies 
is offloaded to the higher stages of the suspension.   
The mirror stage is offloaded to the upper stages at frequencies below \tildegood7~Hz, and the middle stage 
is offloaded to the top at frequencies below \tildegood100~mHz.  

Angular sensing of the relative alignment between the input beam and the IMC is achieved with 
differential wavefront sensing,\cite{anderson_alignment_1984, fritschel_alignment_1998} 
a variant of the PDH technique.  
This technique provides independent error signals for all four relative degrees of freedom and is 
used to force the cavity to follow the input beam with a bandwidth of \tildegood500~mHz.  
In addition, the quadrant detector behind MC2 is used to servo two degrees of freedom of the input 
beam with a bandwidth of \tildegood10~mHz.

%--------------------------------------------------------------------------------------------------
% Auxiliary Mirror Suspensions (HAUXs)
%--------------------------------------------------------------------------------------------------
\subsection{Auxiliary mirror suspensions}

\begin{figure}[t]
\begin{centering}
\includegraphics[width=\linewidth]{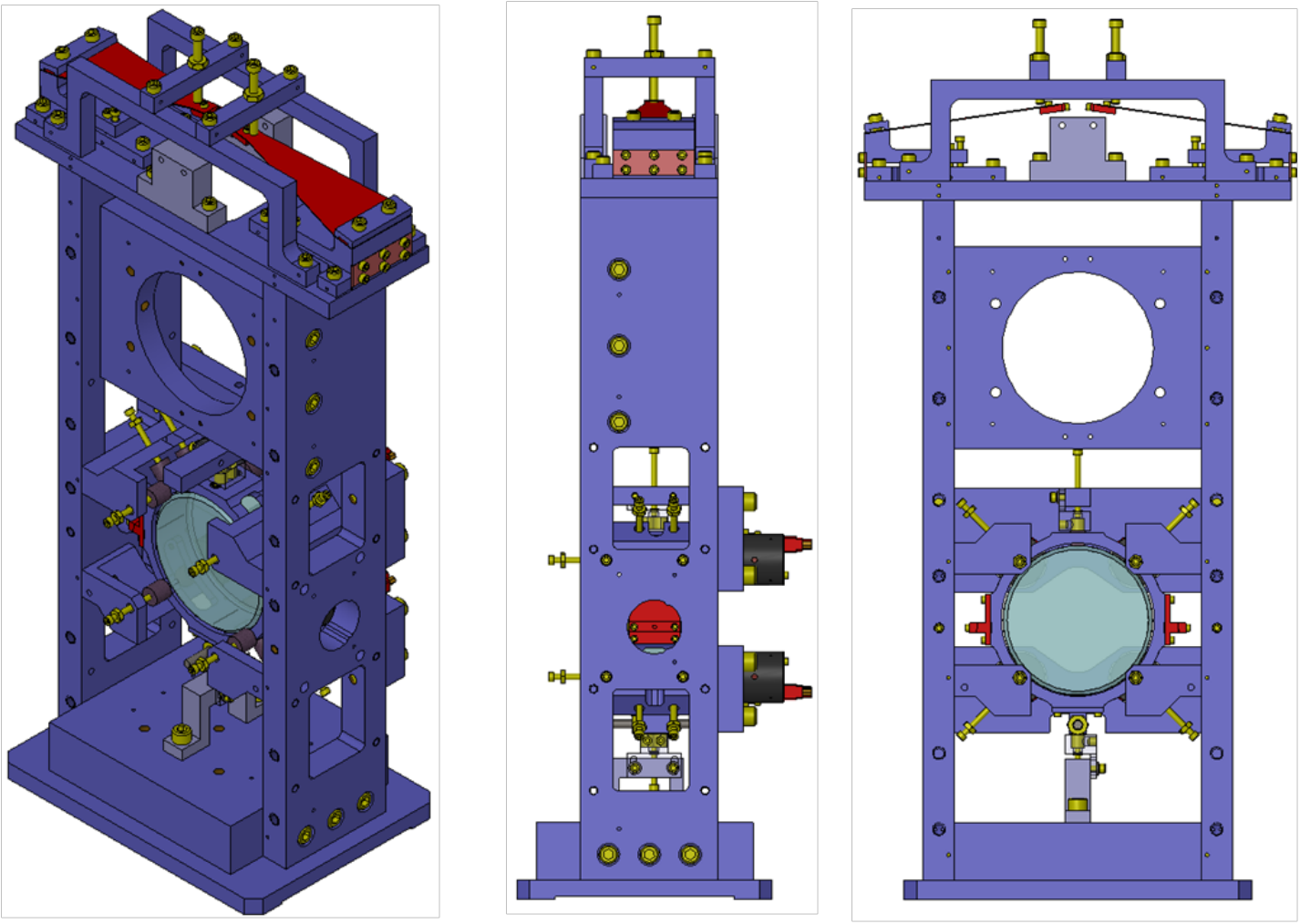}
\par\end{centering}

\protect\caption{\label{fig:The-HAMAux-suspensions} The HAMAux suspensions are
used for the steering mirrors between the mode cleaner and the main
interferometer. These are single stage suspensions with blades for
vertical isolation.}

\end{figure}

The HAM Auxiliary Suspensions (HAUXes), depicted in Figure \ref{fig:The-HAMAux-suspensions}, are 
single pendulum suspensions with the addition of blade springs for vertical isolation. 
The main structure, made of aluminum, fits in an envelope of 127x217x441 mm (DxWxH), weighs 
approximately 6~kg and consists of a base, two side walls, two horizontal bars supporting four
A-OSEMs (a particular variation of the sensors/actuators described in the previous section), a 
stiffening slab connecting the two walls and a top part supporting the blade springs. 
The structure is designed and tested to have the lowest structural resonance above 150 Hz, so as 
not to interfere with the delicate control loops of the LIGO seismic isolation platform on which 
it is installed.  

Two 250~mm long, 150~\textmu m diameter steel music wires run from the tips of two 77~mm long, 
500~\textmu m thick tapered maraging steel blades down to a lightweight circular aluminum holder containing the 
optic. 
The resonant frequencies of the optic's displacement and rotational degrees of freedom are all 
designed to be below 10~Hz which keeps them below the aLIGO measurement band.  
This is particularly important for the degrees of freedom that directly couple into beam motion; 
length, pitch, and yaw.  
According to a numerical model, the pendulum motion around the top suspension point (where the 
wires leave the blades) and the optic pitch motion around the lowest suspension points (where the 
wires attach to the optic holder) combine in two normal modes with frequencies just above and below 
1 Hz.  
The same model shows that the yaw motion around the vertical axis has a slightly lower resonance 
at about 0.8 Hz.

%^^^^^^^^^^^^^^^^^^^^^^^^^^^^^^^^^^^^^^^^^^^^^^^^^^^^^^^^^^^^^^^^^^^^^^^^^^^^^^^^^^^^^^^^^^^^^^^^^^
%^^^^^^^^^^^^^^^^^^^^^^^^^^^^^^^^^^^^^^^^^^^^^^^^^^^^^^^^^^^^^^^^^^^^^^^^^^^^^^^^^^^^^^^^^^^^^^^^^^

The use of an optic holder allows for easy swapping of the optic and
for the use of passive eddy current damping, while providing support for a
balancing threaded rod, to fine tune the optic's pitch, and for the magnets of the 
four \mbox{A-OSEMs}, without the need of gluing them directly to
the optic. The A-OSEMs are arranged in a square pattern thus providing readout and
actuation capabilities for the yaw, pitch and piston degrees of freedom.
The A-OSEMs are used for both beam pointing and local active damping.
Passive eddy current damping along the other degrees of freedom is
provided by two pairs of anti-parallel neodymium magnets attached
to the main structure immediately above and below the optic holder;
their distance from the holder can be changed to tune the damping
action. A set of fourteen soft-tip stoppers protrude from the main
structure towards the optic holder and can be used to mechanically
limit the motion of the optic as well as to securely clamp it during
handling and transportation of the suspensions. 

The HAUXs were assembled on site and characterized prior to installation by acquiring a complete 
set of transfer functions (TFs) from force (torque) applied to the optic to displacement (rotation) 
of the optic. 
For this measurement the A-OSEMs were used as both actuators and sensors. 
It is worth noting that, although the role of the suspension is to isolate the mirror from motion 
of the ground, measurement of TFs from force at the optic to displacement allows to predict the performance of the suspensions.

%^^^^^^^^^^^^^^^^^^^^^^^^^^^^^^^^^^^^^^^^^^^^^^^^^^^^^^^^^^^^^^^^^^^^^^^^^^^^^^^^^^^^^^^^^^^^^^^^^^
\begin{figure}[t]
	\begin{centering}
		\includegraphics[clip,width=\linewidth]{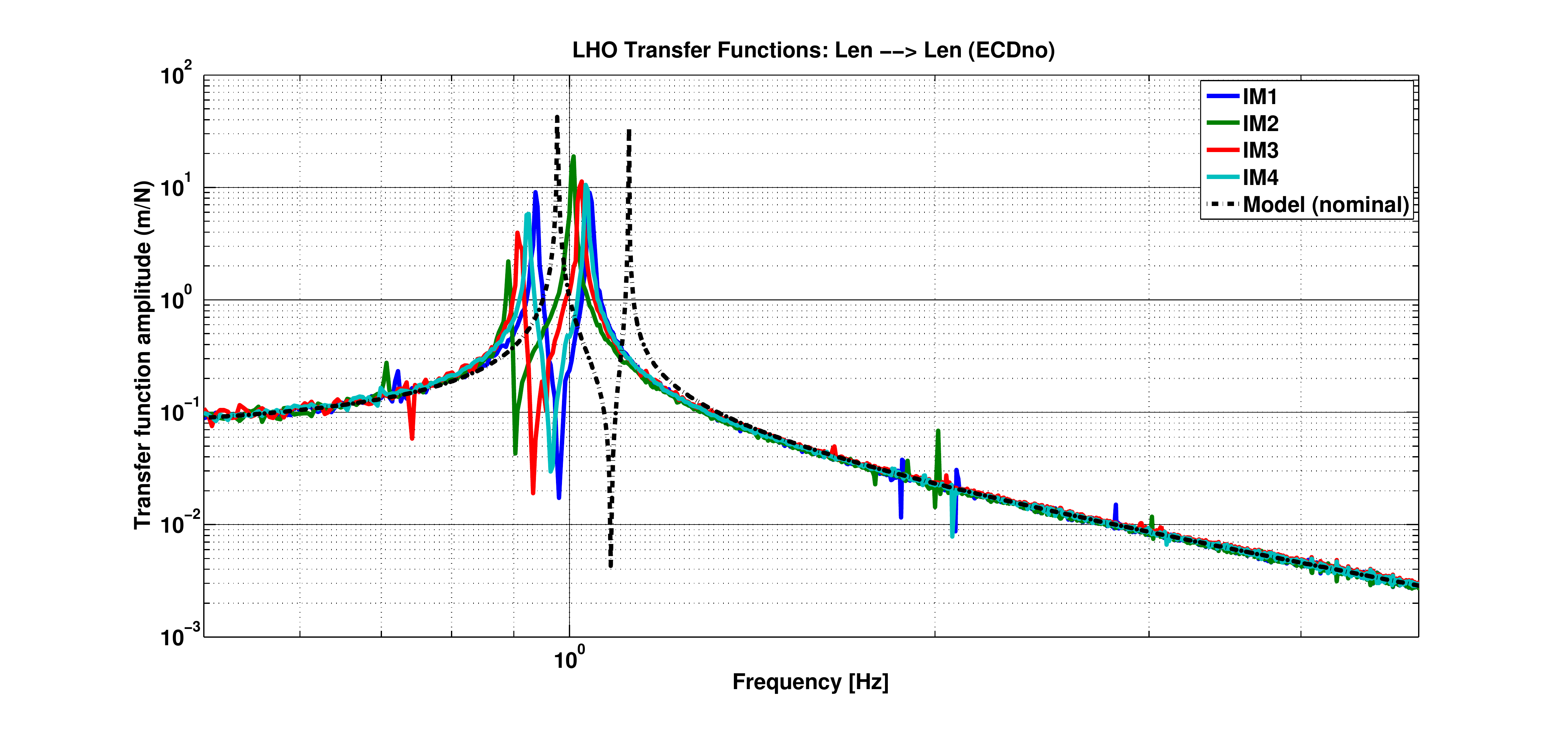}
		\includegraphics[clip,width=\linewidth]{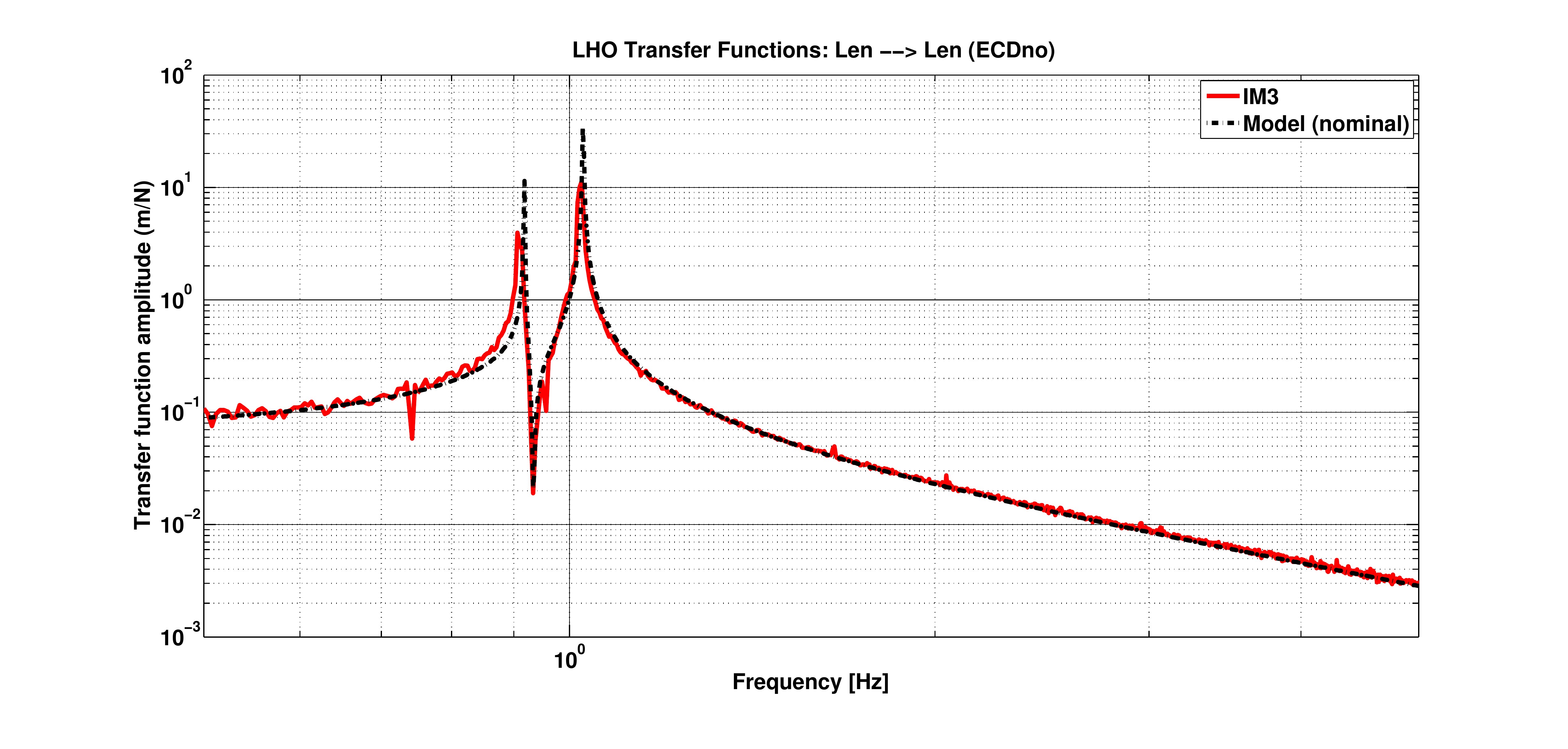}
	\par\end{centering}
	\protect\caption{\label{fig:HAMAux-TFs} 
		Top: the length-to-length transfer functions measured for four HAMAux suspensions, plotted 
		together with the prediction from a numerical model. 
		The differences between nominally identical suspensions are compatible with machining and 
		assembly tolerances. 
		The overall shift of all the measured resonance peaks towards lower frequencies compared to 
		the model is explainable with inaccuracies in some of the parameters used in the model. 
		Bottom: measured and modeled TF can be made to agree very well with realistic variations of 
		the model parameters from their nominal value. Here, the model was specifically fine tuned to 
		match the TF measured on IM3.}
\end{figure}
%^^^^^^^^^^^^^^^^^^^^^^^^^^^^^^^^^^^^^^^^^^^^^^^^^^^^^^^^^^^^^^^^^^^^^^^^^^^^^^^^^^^^^^^^^^^^^^^^^^

TFs were acquired both before and after installation of eddy current dampers, with the former 
providing a better tool to verify overall performance of the suspensions, since the dampers 
broaden the peaks and make it difficult to recognize critical features in the TFs. 
Figure \ref{fig:HAMAux-TFs} shows one example of such a TF (in this case, force versus displacement 
along the beam axis) measured for the 4 different suspensions assembled at LHO, as well as the 
value calculated from the numerical model of the suspensions.

The variability between measured transfer functions, particularly the position of the resonance 
peaks, of nominally identical suspensions is well within what can be explained by machining and 
assembly tolerances. 
Since they are of little consequence, no mechanism has been incorporated in the HAUX design to 
allow for post-assembly compensation of these errors and fine tuning of the resonances.

Of a different nature is the systematic shift of the resonant frequencies of all measured TFs 
(including the ones not shown here) towards lower values compared to those predicted. 
It is possible to explain the observed discrepancy by assuming that the actual value of some of 
the parameters used in the numerical model, e.g. material properties or quantities estimated from 
the CAD drawing, is different from the nominal one. 
Unfortunately, this procedure yields multiple possible combinations of realistic 
parameter values; the result obtained from one particular choice of parameters is shown 
in the bottom of Figure \ref{fig:HAMAux-TFs}. 
Identifying the correct value of the parameters would have required a long dedicated 
measurement campaign; given the time constraints imposed by the aLIGO installation schedule and 
the fact that a shift of resonances towards lower frequencies is actually an advantage, this was 
deemed unnecessary.

Regardless of the previous two observations, the transfer functions show the expected overall 
behavior and isolation in the measurement band, and the resonances are all below 10 Hz; the latter 
is true also for the degrees of freedom that are not nominally sensed by the A-OSEMs, but whose 
resonances are nevertheless detectable because of cross couplings into the A-OSEMs readout. 
The HAM Auxiliary Suspensions thus satisfy the design requirements.

The complete sets of transfer functions to length, pitch and yaw as well as the cross couplings 
have been measured for all HAUXs and are documented under LIGO document number 
T1300030.\cite{dcc_t1300030} 
All design drawings, materials, assembly instructions, and test protocols and many other documents 
related to the HAUX suspensions are available under LIGO document number 
T1300020.\cite{dcc_t1300020}

%--------------------------------------------------------------------------------------------------
% Mode Matching from the PSL to the Main Interferometer
%--------------------------------------------------------------------------------------------------
\subsection{Mode matching from the PSL to the main interferometer}

In order to couple the maximum amount of light from the PSL into the interferometer, it is 
important that the spatial mode of the laser beam be properly matched to the spatial mode of the 
resonant interferometers downstream.  
The input optics is responsible for matching the mode from the PSL into the IMC as well as 
matching the mode from the IMC into the interferometer.  
In both cases Galilean telescopes are used to avoid tightly focused beams and shorten the 
telescope lengths.  

The mode coming from the PSL is defined by the pre-mode cleaner (PMC, Figure \ref{fig:PSL}) which has 
a waist size of 550~$\mu$m.  
This is mode matched to the IMC by a pair of 50~mm diameter lenses rigidly mounted on the PSL table 
between the EOM and the power control stage.  
The first lens has a focal length of $-459.5$~mm, the second lens has a focal length of 
$+1145.6$~mm, the separation between the two is 838~mm, and the first lens is placed 890~mm from 
the PMC waist.  
The total distance from the PMC waist to the HR surface of MC1 is roughly 10.9~m.
The IMC mode has a waist located midway between MC1 and MC3 (Figure \ref{fig:The-in-vacuum-parts}) 
with a size of 2.12~mm.  
The spatial mode of the IMC is mode matched to the main interferometer by the suspended mirrors 
IM2 and IM3 (Figure \ref{fig:The-in-vacuum-parts}) which sit before and after the FI respectively.  
The focal length of IM2 is 12.8 m, the focal length of IM3 is -6.24 m, and the separation between 
them is 1.170 m.  
IM2 is located 1.78~m downstream of the HR surface of MC3, and the total distance from the HR 
surface of MC3 to that of PRM (Figure \ref{fig:IFO}) is 4.59~m.

A detailed as-built layout (to scale) of the PSL table, including the relevant components of the 
input optics can be found at LIGO document number D1300347\cite{dcc_d1300347} for LLO and 
D1300348\cite{dcc_d1300348} for LHO.  
The coordinates of all in-vacuum optics can be found in the master coordinate list under LIGO 
document number E1200274\cite{dcc_e1200274} for LLO and E1200616\cite{dcc_e1200616} for LHO.

%==================================================================================================
%  Performance of the Input Optics
%==================================================================================================
\section{Performance of Input Optics }
\label{sec:performance}

This section discusses the integrated tests performed on the input optics after everything was 
installed into the vacuum system.  
Although the output power of the PSL is capable of reaching $180\ \text{W}$, LIGO has chosen for 
technical reasons to operate at powers below $30\ \text{W}$ for the foreseeable future.  
This limitation prevented extended, in-vacuum tests of the FI and IMC at the high powers for which 
they were designed.  

%************************************************
% Paragraph: Power Budget
%************************************************
\paragraph*{Power Budget}

A power budget of the input optics (IO) was made at both sites using a calibrated power meter to measure 
optical powers at various points throughout the system.  
Optical power at various key points in the vacuum system was inferred from this data by using the 
expected transmissivities of the pickoff mirrors, and power coupling between these key points was 
inferred from this data.  
Table \ref{tab:performance_power_budget} shows the calculated power coupling between the various 
key points of the IO for LLO (results are similar for LHO).  
Note that the overall IO efficiency of 84.5~\% does not include the mode mismatch into the 
interferometer.  
Measurement of the mode mismatch is complicated since it competes with the mode matching between the power recycling cavity and the arm cavities as well as the impedance of the 
interferometer which, in turn, depends on knowing the precise reflectivity of all of the mirrors.  
The reflected light from the interferometer when it is held on resonance sets an upper limit on 
the mode mismatch of \tildegood9\%.

%^^ Table: Power Budget ^^^^^^^^^^^^^^^^^^^^^^^^^^^^^^^^^^^^^^^^^^^^^^^^^^^^^^^^^^^^^^^^^^^^^^^^^^^
\begin{table}[t]
	\centering
	\renewcommand{\arraystretch}{1.5}
	\begin{tabular}{lc}
		\hline 
		Path                & Power Coupling (\%) \\
		\hline 
		PSL to MC1          & $95.3\pm1.3$ \\
		IMC visibility      & $98.4\pm0.1$ \\
		IMC transmissivity  & $92.2\pm3.1$ \\
		MC3 to PRM	        & $97.9\pm2.9$ \\
		Full IO: PSL to PRM & $84.5\pm2.5$ \\ 
		\hline
	\end{tabular}
	\caption{The power coupling between various points in the input optics.  
		The overall transmissivity of the IO does not include mode matching losses into the main interferometer.}
	\label{tab:performance_power_budget}
\end{table}		
%^^^^^^^^^^^^^^^^^^^^^^^^^^^^^^^^^^^^^^^^^^^^^^^^^^^^^^^^^^^^^^^^^^^^^^^^^^^^^^^^^^^^^^^^^^^^^^^^^^

%************************************************
% Paragraph: In-vacuum FI Isolation Ratio
%************************************************
\paragraph*{In-vacuum FI Isolation Ratio:} 
The in-vacuum isolation performance of the FI was measured at low power.  
The measurement was made by placing a pair of matched beamsplitters into the laser beam on the PSL 
table between lenses L1 and L2 (see Figure \ref{fig:PSL-IO}), each having a reflectivity of $32\ \%$, 
and a photodiode was added to the backwards propagating beam from the IFO.  
With the IMC locked, the angle of PRM was adjusted in order to maximize the power on the PD.  
Taking into account the losses from mode matching to the IMC and the reflectivity of PRM, 
an isolation ratio of $29.1\ \text{dB}$ was measured at LLO and $35.0\ \text{dB}$ at LHO.  
Note that the HWP angle inside the FI was not adjusted to optimize the isolation ratio. The measured isolation ratios appear to be sufficient and rotating the HWP in situ was seen as an unnecessary risk.

%************************************************
% Paragraph: IMC Cavity Pole
%************************************************
\paragraph*{IMC Cavity Pole:} 
The IMC, like all optical cavities, acts as a low pass filter to variations in both laser 
frequency and intensity.  
The $-3\ \text{dB}$ point of this low pass filter, the so called cavity pole, is a function of the 
reflectivity of the mirrors as well as the round-trip losses;
\begin{equation}
	\Omega_0=\frac{c}{L}\frac{1-r^2\sqrt{1-\frac{\ell}{r^2}}}{r^2\sqrt{1-\frac{\ell}{r^2}}},
	\label{eq:performance_cavity_pole_w_losses}
\end{equation}
where $r$ is the amplitude reflectivity of MC1 and MC3 (assumed equal) and $\ell$ is the 
round-trip loss.  

\begin{figure}[t]
	\centering
	\includegraphics[width=\linewidth]{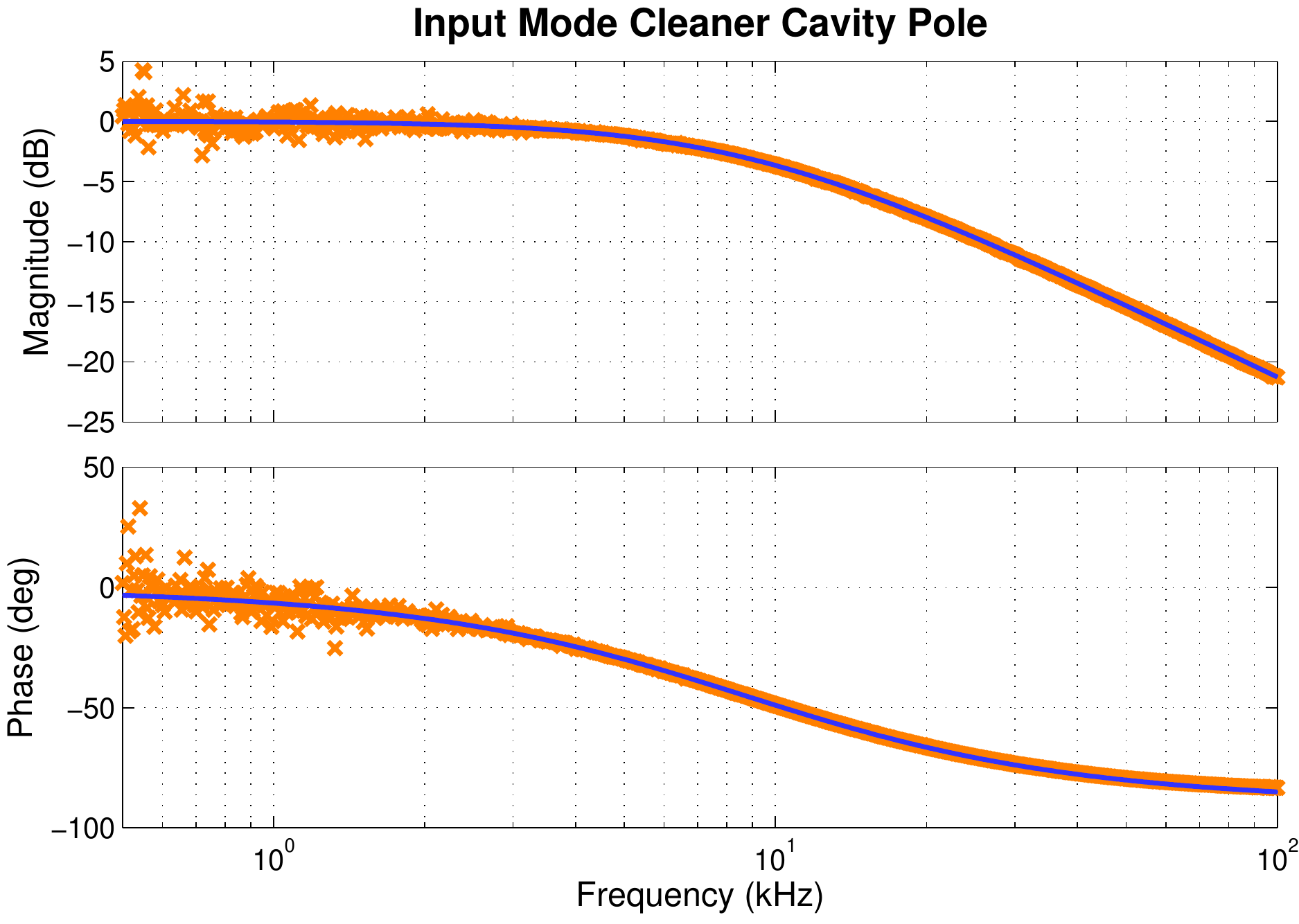}
	\caption{The measured IMC cavity pole is shown together with a simple single pole fit.}
	\label{fig:performance_imc_cavity_pole}
\end{figure}

The cavity pole of the IMC was measured by amplitude modulating the input beam and taking the 
transfer function between an RFPD before the IMC and another RFPD after the IMC.  
Figure \ref{fig:performance_imc_cavity_pole} shows the data taken at LLO together with a single pole fit 
to the data.  
The fit has a pole frequency of $8,686\pm108\ \text{Hz}$ which gives an IMC finesse of $522$.  
Using the vendor measured transmissivity of MC1 and MC3, this gives a round-trip loss via 
\eqref{eq:performance_cavity_pole_w_losses} of $\ell=164\pm147\ \text{ppm}$.  

%^^^^^^^^^^^^^^^^^^^^^^^^^^^^^^^^^^^^^^^^^^^^^^^^^^^^^^^^^^^^^^^^^^^^^^^^^^^^^^^^^^^^^^^^^^^^^^^^^^
%^^^^^^^^^^^^^^^^^^^^^^^^^^^^^^^^^^^^^^^^^^^^^^^^^^^^^^^^^^^^^^^^^^^^^^^^^^^^^^^^^^^^^^^^^^^^^^^^^^

%************************************************
% Paragraph: IMC Scattering
%************************************************

\paragraph*{IMC Scattering:}
Scattering from the IMC mirrors was measured in-situ at LLO using calibrated digital (GigE) cameras.  
The available views of the optics were restricted by the vacuum system to five different locations 
which gives seven different views of the three optics.  
The extra two views are due to the fact that the scattering from MC1 and MC3 can be seen in the 
reflection of each other for angles near the beam line.  
The cameras were calibrated with an $1.064\ \mu\text{m}$ laser source for various 
optical powers and camera exposure/gain settings.  

The surface roughness of the IMC mirrors is required to have an RMS deviation below 0.1~nm.  
This puts the mirrors of the IMC into the smooth surface regime in which the fluctuations of the 
surface height are significantly less than the wavelength of the light.  
This regime of optical scattering is governed by the Rayleigh-Rice 
theory\cite{schroder_modeling_2011} in which the angular distribution of the scattered light is 
governed solely by the statistical properties of the surface height 
fluctuations.\cite{church_relationship_1979}
In particular, the angular distribution of the scattered light is determined by a simple mapping 
from the two-dimensional power spectral density of the surface height variations.  
Each spatial wavelength can be thought of as a diffraction grating which contributes to the 
scattering at the first order deflection given by
\begin{equation}
	\sin\theta_s=\sin\theta_i\pm\frac{\lambda}{d},
\end{equation}
where $d$ is the spatial wavelength of interest.

%^^ IMC: Scatterng Results Table ^^^^^^^^^^^^^^^^^^^^^^^^^^^^^^^^^^^^^^^^^^^^^^^^^^^^^^^^^^^^^^^^^^
\begin{table}[t]
	\centering
	\renewcommand{\arraystretch}{1.5}
	\begin{tabular}{ccccc}
		\hline
		Optic & $\theta_i$ (deg) & $\theta_s$ (deg) & 
			BRDF $(\tfrac{10^{-6}}{\text{sr}})$ & $d\ (\mu\text{m})$  \\
		\hline
		MC1 & $44.6$ & $45.9$	 & $153\pm5$       & $66.6$ \\
		MC1 & $44.6$ & $22.5$  & $0.5\pm0.1$     & $3.3$  \\
		MC1 & $44.6$ & $-42.8$ & $0.025\pm0.001$ & $0.8$  \\
		MC2 & $0.8$  & $2.0$   & $5569\pm71$     & $50.8$ \\
		MC2 & $0.8$  & $61.2$  & $8.9\pm1.9$     & $1.2$  \\
		MC3 & $44.6$ & $45.9$  & $95.7\pm4.6$    & $66.6$ \\
		MC3 & $44.6$ & $-42.8$ & $0.102\pm0.002$ & $0.8$  \\
		\hline
	\end{tabular}
	\caption{The scattering results for the three mirrors of the input mode cleaner.  
		The incident and scattering angles, measured with respect to the optic normal are shown 
		together with the measured BRDF and the spatial wavelength which leads to diffraction at the 
		scattering angle.}  
	\label{tab:imc_scatter_results}
\end{table}
%^^^^^^^^^^^^^^^^^^^^^^^^^^^^^^^^^^^^^^^^^^^^^^^^^^^^^^^^^^^^^^^^^^^^^^^^^^^^^^^^^^^^^^^^^^^^^^^^^^

Table \ref{tab:imc_scatter_results} shows the results of these scatter measurements.  
The angle of incidence, determined by the IMC geometry, and the angle of scattering, determined by 
the available view of the optic, taken with respect to the optic normal, are shown in the first 
two columns of the table.  
The second two columns show the measured bi-directional reflectance distribution function (BRDF) 
and the spatial wavelength of the surface 
deformations which lead to scattering at that angle within the Rayleigh-Rice formalism.  
While the surface scatter of MC1 and MC3 are roughly as expected, the surface scatter of MC2 
is high by more than an order of magnitude.  

An estimate of the total integrated scatter (TIS) was made for all three optics by first fitting 
the measured data to the function $A/(\theta_s-\theta_i)^2$, where $A$ is the fitting parameter.  
This function is empirically motivated and is typical of surface scattering seen from high 
quality laser line optics.\cite{magana-sandoval_large-angle_2012, gloge_scattering_1969}
Integrating this function over azimuthal angles from $\pi/2$ down to the beam divergence angle 
gives an estimate of the total scatter of light out of the beam.  
Doing so gives TIS values of 3.8 ppm for MC1, 90.2 ppm for MC2, and 5.7 ppm for MC3.

%************************************************
% Paragraph: IMC Absorption
%************************************************
\paragraph*{IMC Absorption:} 
The total absorption of the mirrors in the IMC at LLO was measured by tracking the frequency of the 
resonance of the TEM$_{01}$ mode while the circulating power was intermittently increased and decreased.  
For an ideal cavity with spherical mirrors, the round-trip Gouy phase of the cavity is determined 
entirely by the radii of curvature of the mirrors and the distances between them.  
The location of the TEM$_{01}$ resonance is determined by this round-trip Gouy phase, and 
measurement of its location therefore provides a precise method for measuring changes in the radii 
of curvature of the mirrors.  
To first order a mirror which is heated by optical absorption deforms 
spherically\cite{winkler_heating_1991}.  
Tracking of the location of this resonance while modulating the power therefore provides a method 
of measuring the absorption of the mirrors.  

The location of the TEM$_{01}$ resonance was measured by driving the EOM with the RF output of a 
network analyzer, adding phase sidebands to the carrier beam which was held on resonance by the 
control system of the IMC.  
The signal of an RF photodiode in transmission of the IMC was demodulated by the network analyzer 
so that, when the frequency swept across the TEM$_{01}$ resonance, the 
beating between the sideband and the carrier mapped out the resonance of the first order mode.  
In addition, offsets were inserted into the IMC angular control loops in order to keep it slightly 
misaligned and enhance the relative TEM$_{01}$ content of the sideband, and a small portion of the transmitted beam was 
occluded before being sent to the RFPD to enhance the beat signal between the fundamental mode of 
the carrier and the TEM$_{01}$ mode of the sideband.  
This setup was used to monitor the location of the first order resonance of the IMC while the 
input power was intermittently increased and decreased.  
Further information about the details of this technique can be found in Reference 
\onlinecite{mueller_in-situ_2015}.

%^^ IMC: Losses: Absorption Measurement Results ^^^^^^^^^^^^^^^^^^^^^^^^^^^^^^^^^^^^^^^^^^^^^^^^^^^
\begin{table*}[t]
	\centering
	\renewcommand{\arraystretch}{1.5}
	\begin{tabular}{ccccc}
		\hline
		Date &  Power (W)  &  $f_{01}$ (Hz) & 	$\Delta f_{01}$ (Hz) & 
			  Abs. (ppm/mir.)  \\
		\hline
		\multirow{2}{*}{1/17/2013} & 	3.11 & $29,266,891\pm60$ & \multirow{2}{*}{$6,230\pm69$} 
			& \multirow{2}{*}{$2.39\pm0.02$}\\
		& 30.5 & $29,273,121\pm35$ & & \\
		\hline
		\multirow{2}{*}{7/23/2013} & 	0.517 & $29,268,352\pm65$ & \multirow{2}{*}{$245\pm85$} 
			& \multirow{2}{*}{$1.50\pm0.46$}\\
		& 1.034 & $29,268,597\pm56$ & & \\
		\hline
		\multirow{2}{*}{8/6/2013} & 	0.203 & $29,267,831\pm14$ & \multirow{2}{*}{$358\pm15$} 
			& \multirow{2}{*}{$1.42\pm0.06$}\\
		& 1.011 & $29,268,189\pm6.4$ & & 	\\
		\hline
		\multirow{2}{*}{9/30/2013} & 	1.074 & $29,266,056\pm39$ & \multirow{2}{*}{$1,175\pm54$} 
			& \multirow{2}{*}{$1.53\pm0.06$}\\
		& 3.076 & $29,267,230\pm37$ & & \\
		\hline
		\multirow{2}{*}{6/17/2014} & 1.79 & $29,877,736\pm46$ & \multirow{2}{*}{$2,249\pm79$}
			& \multirow{2}{*}{$0.50\pm0.01$}\\
		& 10.24 & $29,875,487\pm64$ & & \\
		\hline
	\end{tabular}
	\caption{The data and inferred absorption from numerous repetitions of the Gouy phase absorption 
		measurements in the IMC.
		The Power and $f_{01}$ columns show the input power level and location of the TEM$_{01}$ peak 
		while the $\Delta f_{01}$ and Abs. columns show the shift in this peak between power levels 
		and the inferred absorption.} 
	\label{tab:imc_gouy_phase_absorption}
\end{table*}
%^^^^^^^^^^^^^^^^^^^^^^^^^^^^^^^^^^^^^^^^^^^^^^^^^^^^^^^^^^^^^^^^^^^^^^^^^^^^^^^^^^^^^^^^^^^^^^^^^^

Table \ref{tab:imc_gouy_phase_absorption} shows the results of several repetitions of this 
measurement at LLO over the course of nearly one and a half years.  
The second column shows the two power levels which were used for each measurement, the third 
column shows the measured frequencies at those power levels, and the fourth column shows the 
shift in the frequencies.  
The last column shows the amount of inferred absorption per mirror based on a numerical model 
developed using a finite element simulation to calculate the thermal deformation and an FFT based 
beam propagation simulation to calculate the shift of the resonance.   
Although the absorption is slightly higher than was anticipated, the amount of inferred thermal 
lensing from absorption at this level only leads to a $0.3\ \%$ reduction in power coupled into 
the interferometer.

%************************************************
% Paragraph: IMC Length
%************************************************
\paragraph*{IMC Length}

The same setup used to measure the IMC absorption was also employed to measure the length of the 
IMC.  
By adding a small offset to the error point of the control loops which hold the IMC on resonance, 
the phase modulation (PM) impressed by the network analyzer via the EOM gets converted to amplitude modulation 
(AM) on either side of the cavity's free spectral range.  
Precisely on resonance the magnitude of this PM to AM conversion goes through a minimum, and the 
phase flips sign.  
Note that this is the same effect on which the Pound-Drever-Hall technique for cavity locking is 
based\cite{drever_laser_1983}.  
By scanning the sidebands across successive free spectral ranges the length of the IMC was 
measured at LLO to be $32,947.3\pm0.1$~mm and at LHO to be $32,946.6\pm0.1$~mm

%************************************************
% Paragraph: IMC Noise Budget
%************************************************
\paragraph*{IMC Noise Budget:}

As with all lock-in experiments, the feedback signals to the length and frequency paths of the 
IMC control system are a measure of the fluctuations in these quantities.  
Understanding the source of these fluctuations is important because some noise sources will be 
suppressed either actively or passively while others, e.g. sensing noises, can be impressed by the 
control system and inject noise into the main interferometer.  

%^^ IMC: Noise: Noise Budget ^^^^^^^^^^^^^^^^^^^^^^^^^^^^^^^^^^^^^^^^^^^^^^^^^^^^^^^^^^^^^^^^^^^^^^
\begin{figure}[t]
	\centering
	\includegraphics[width=\linewidth]{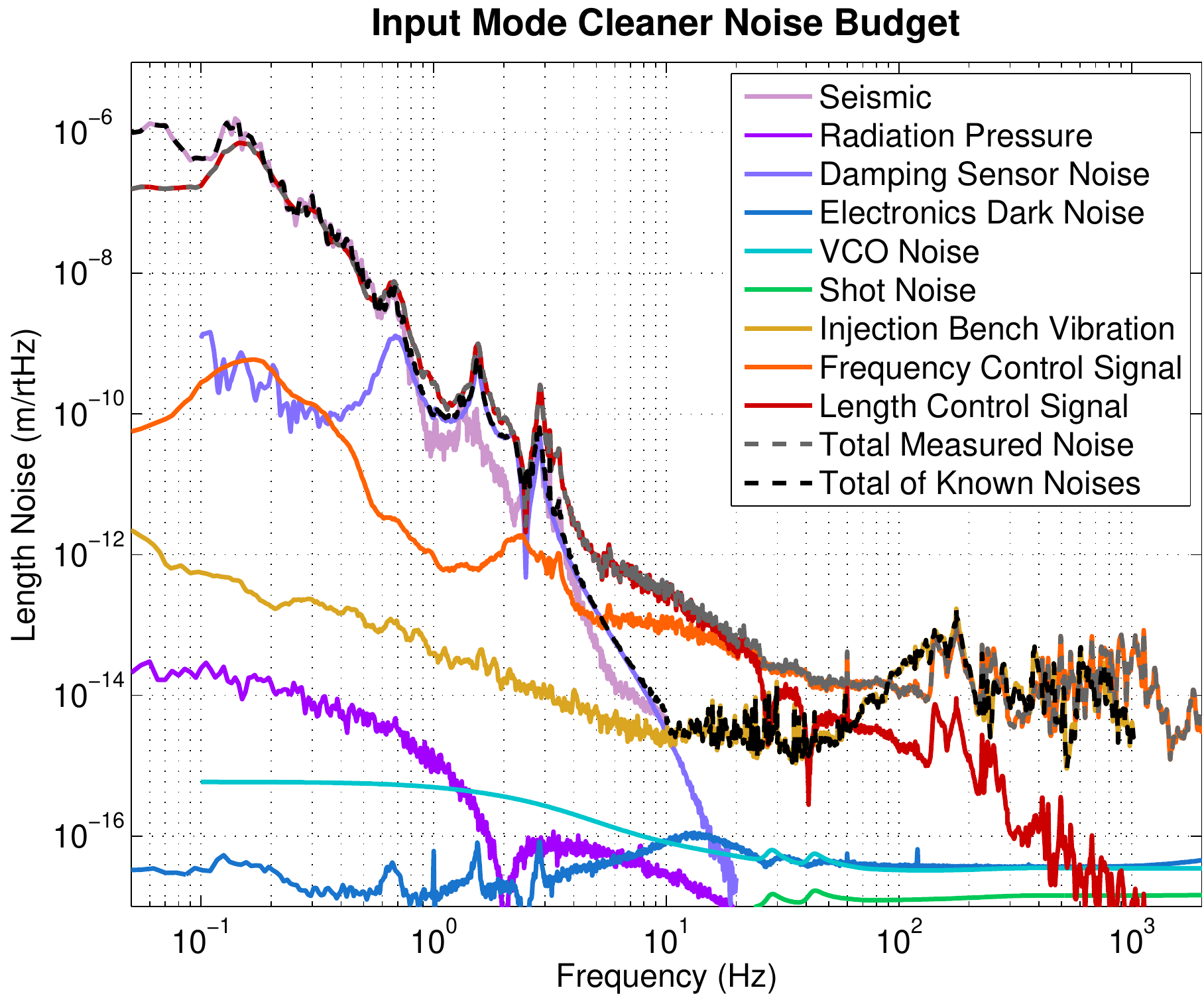}
	\caption{The noise budget of the aLIGO input mode cleaner.  
		The measured noise in the length path is shown in red, the measured noise of the frequency 
		path is shown in orange, and the sum total of these two is shown in dashed gray.  
		The sum total of the understood noises is shown in dashed black, and the individual terms are 
		discussed in the text.}
	\label{fig:imc_noise_budget}
\end{figure}
%^^^^^^^^^^^^^^^^^^^^^^^^^^^^^^^^^^^^^^^^^^^^^^^^^^^^^^^^^^^^^^^^^^^^^^^^^^^^^^^^^^^^^^^^^^^^^^^^^^

The noise budget for the IMC is shown in Figure \ref{fig:imc_noise_budget}.  
The red and orange curves show the feedback signals to the length and frequency respectively, and 
the dashed gray line shows the coherent sum of these two.  
The other curves on the plot are a mixture of measured noise and theoretical predictions 
propagated to the error point with a controls model.  
At low frequencies, below \tildegood5~Hz, the dominant sources of noise are from seismic motion 
of the mirror suspensions and sensor noise injected through the active control loops.  
At high frequencies, above \tildegood80~Hz, the dominant noise is caused by frequency noise of the
incident laser beam which is created, at least in part, by vibrations of the injection 
bench on which the PSL is built (see yellow and orange trace in Figure \ref{fig:imc_noise_budget}).

The missing noise between $5\ \text{Hz}$ and $80\ \text{Hz}$ was the subject of many 
investigations, but is still not fully understood.  
The level of the noise in this region averages around that shown in Figure 
\ref{fig:imc_noise_budget} but fluctuates on day timescales by as much as an order of 
magnitude but never coming down to the level of the understood noise.

%==================================================================================================
%  Outlook and Conclusions
%==================================================================================================
\section{Outlook and conclusions}
\label{sec:conclusion}

In summary, we have presented the design of the major components of the aLIGO input optics in full 
detail.  
Our EOM is capable of simultaneously adding three RF sidebands with minimal 
thermal lensing at CW powers of up to 200 W all while maintaining an RFAM level below $10^{-4}$.  
Our FI design provides greater than 30 dB of isolation at CW powers of up to 200 W 
(single pass) with relatively small thermal lensing.  
The input mode cleaner uses mirrors hanging from the aLIGO triple suspensions to provide 
sensing limited length noise fluctuations of \tildegood$1\ \frac{\text{fm}}{\sqrt{\text{Hz}}}$ 
above 10 Hz while transmitting 165 W of CW laser power.  
Finally, the HAM auxiliary suspensions isolate the laser beam from seismic fluctuations above 
their pole frequencies near 1~Hz while allowing for active control of the pointing of the laser 
beam into the interferometer.  

We also presented a comprehensive set of tests of the full integrated input optics system
which showed that the individual components of the input optics continue to function well when 
integrated into the aLIGO interferometers.  
We showed that the full integrated system is capable of delivering 84.5 \% of the incident power 
to the input of the main interferometer.  
We showed that the FI provide greater than 29 dB of isolation at low power, a 
value which is expected to rise once the half-wave plate has been optimized.  
The scattered light from the optics of the IMC was measured and found to be below 6 ppm except 
for one mirror whose scattering is expected to come down with a subsequent cleaning.  
The absorption of the IMC was also measured to be consistently below 5 ppm per mirror over the 
course of nearly one and a half years.  
A noise budget for the IMC was presented which showed that the IMC is frequency noise limited 
above 100 Hz and can therefore serve as an reference for active control of the incident laser 
frequency variations.

\bibliography{iobib}

\end{document}